\documentclass{aa}  
\usepackage{amsmath,amsfonts,amssymb}
\usepackage{textcomp,gensymb}
\usepackage[T1]{fontenc}
\usepackage{aecompl}
\usepackage{ulem}
\usepackage{graphicx}
\usepackage[usenames]{xcolor}
\usepackage{color}
\usepackage{times}
\usepackage{subfig}
\usepackage{stfloats}
\usepackage[utf8]{inputenc}
\usepackage{hyperref}
\usepackage{multirow}
\usepackage{enumerate}
\usepackage{physics}
\usepackage{chemarrow}
\usepackage{tikz}
\usepackage{txfonts}
\usepackage{booktabs}
\usepackage[flushleft]{threeparttable}

\usetikzlibrary{shapes,arrows}
\usetikzlibrary{shapes.geometric}

\newcommand{\rev}[1]{\textcolor{black}{#1}}
\newcommand{\revII}[1]{\textcolor{black}{#1}}

\begin{document} 


   \title{The HR 8799 Debris Disc: Shaped by Planetary Migration and a Possible Fifth Outermost Planet}

   \subtitle{}

   \author{Pedro P. Poblete\inst{1},
          Tim D. Pearce\inst{2}
          and
          Carolina Charalambous\inst{3}
          }

   \institute{Univ. Grenoble Alpes, CNRS, IPAG, 38000 Grenoble, France\\
              \email{pedro.poblete@univ-grenoble-alpes.fr}
    \and
        Department of Physics, University of Warwick, Coventry CV4 7AL, UK
    \and
        Instituto de Astrofísica, Pontificia Universidad Católica de Chile, Av. Vicuña Mackenna 4860, 782-0436 Macul, Santiago, Chile 
    }

   \date{Received January XX, XXXX; accepted March XX, XXXX}

\titlerunning{Exploring HR~8799 with outward migration}

 
  \abstract
   {The HR~8799 system, hosting four giant planets between a warm and cold debris disc, and an extended dusty tail beyond, serves as an ideal laboratory for studying planetary formation and evolution. The debris discs have been observed across various wavelengths, and the planetary properties are well-constrained. Nonetheless, there are still open questions regarding the role of the planets in shaping the debris discs.}
   {We investigate the system's evolution with the aim of understanding how planetary migration shapes its architecture, both in terms of planets and the disc.}
   {We performed $N$-body simulations to model the HR~8799 system. We examined the orbital evolution of the known four super-Jupiter planets through the course of \rev{simple, imposed} migration in a gaseous \rev{disc} as they perturb an external, massless planetesimal disc. We also explore the impact of introducing a fifth planet on the dynamical and morphological aspects of the disc.}
   {The planets migrate outward as a result of their \rev{imposed} interactions with the gaseous disc while maintaining their resonant configuration. This outward migration excites the planetesimal disc, producing a transient scattered population. While a four-planet system partially reproduces the observed cavity between the star and the cold debris disc, the inclusion of a fifth low-mass planet appears to be crucial for better reproducing key morphological aspects of the cold debris disc.}
   {This model provides a novel explanation for the architecture of HR~8799. Outward planetary migration, combined with mean motion resonant interactions and a fifth, low-mass planet, can effectively replicate the observed planetary architecture and cold debris disc characteristics. Our findings underscore the \rev{potential} important role of planetary migration in shaping debris discs.}

   \keywords{Interplanetary medium --- Celestial mechanics --- methods: numerical ---  planets and satellites: fundamental parameters --- planet-disc interactions --- planets and satellites: dynamical evolution and stability}

   \maketitle
%


\label{firstpage}


\section{Introduction}
\label{sec:intro}

The architecture of a planetary system emerges from the complex interplay of various processes, shaping its final architecture. This begins with the formation of a protoplanetary disc (PPD) around a central star, where dust particles coagulate into planetesimals \citep[see][for a comprehensive review]{Armitage2018}. In this environment, protoplanets emerge and, in the process, perturb their surroundings. Protoplanets acquire most of their properties from the PPD, such as mass and chemical composition; however, their final positions are also influenced by interactions with other protoplanets and the gaseous disc, particularly during the final stages of formation. Understanding this stage is crucial for reconstructing the evolutionary paths of planets from their formation to their currently observed positions.

The most extensively studied planetary system is, without a doubt, our Solar System. The abundance of available data provides strong constraints on models aiming to explain its architecture, particularly concerning planetary orbital parameters, and the characteristics of the asteroid and Kuiper belts. One prominent model for the Solar System's evolution is the Nice model, which describes the current planetary locations while also reproducing planetesimal populations in the Kuiper belt \citep{Gomes+2005, Tsiganis+2005, Morbidelli+2005, Levison+2008, Nesvorny+2012}. This early instability is thought to arise from a combination of inward and outward migration of the planets, possibly occurring while gas was still present in the Solar protoplanetary disc \citep{Griveaud+2024}. Thus, the Nice model offers a consistent and plausible evolutionary path for the Solar System. Then, a natural question arises as to whether other planetary systems might have undergone similar evolutionary processes.

The search for and characterization of exo-planets is a highly active field. Since the first discovery of an exoplanet orbiting a main sequence star, 51~Pegasi b \citep{Mayor+1995}, the catalog of exoplanets has grown substantially. Detection capabilities were significantly enhanced with the Kepler mission \citep{Borucki+2010} and later with the TESS mission \citep{Ricker+2015}. Currently, there are approximately 7,000 confirmed exoplanets, with about 1,000 in multi-planet systems\footnote{Taken from the encyclopedia of exoplanetary systems \citet{Schneider+2011}, website: \hyperlink{https://exoplanet.eu/catalog/}{https://exoplanet.eu/catalog/}.}. Such multi-planet systems provide a unique opportunity to study the evolution of planetary architectures. In this context, the HR 8799 system stands out as a prime candidate for such studies. Its unique combination of four giant planets with well-characterized orbits, along with well-studied debris discs, makes it an ideal candidate for probing the dynamical evolution of an extrasolar system.


In this study, we explore a novel planetary evolutionary scenario that deviates from classical migration models.  Specifically, we investigate the orbital evolution of super-Jupiter planets undergoing outward and inward migration phases.
We perform a parametric analysis to explore evolutionary pathways that \rev{could} lead to the current observed architecture of the HR 8799 system. In doing so, we focus on the planets' semi-major axes, eccentricities, and resonant angles of the planets, along with the planetesimal distribution within the debris disc. The remainder of the paper is organized as follows: Section \ref{sec:HR} provides a summary of the properties, observations, and previous models of the HR~8799 system; Section \ref{sec:methods} details the numerical techniques and the employed migration prescriptions; Section \ref{sec:results} presents the main results, which are further discussed in Section \ref{sec:discussion}. Finally, conclusions are drawn in Section \ref{sec:conclusion}.

\section{The HR~8799 system}\label{sec:HR}

The HR~8799 (HD 218396) system is an exemplary case of exoplanets detected through direct imaging. It has been extensively studied and is often used as a model laboratory for planetary formation theories. The system is located approximately 41 parsecs away \citep{Gaia+2021} and consists of a central star, four directly imaged planets, and multiple debris populations, including a well-characterized debris disc exterior to the known planets. Below, we describe the system in detail.

\subsection{Architecture}
\label{sec:Architecture}
The central star, with a dymamical mass of $1.516^{+0.038}_{-0.024}\ M_\odot$ \citet{Baines+2012}, is classified as F0V/kA5mA5 \citep{Gray+2003}. 
It is a variable star of the $\gamma$~Doradus type \citep{Zerbi+1999,Kaye+1999} and is identified as a $\lambda$~Bo{\"o}tis star due to its unique chemical abundance \citep{Gray+1999,Sadakane2006,Moya+2010a}. Its age, however,  is uncertain, with values ranging from 25–60 Myr \citep{Zuckerman&Song2012,Rhee+2007,Bell+2015} to as much as 1 Gyr \citep{Moya+2010b}, though the consensus generally lies within the range of 30–40 Myr. 

Four super-Jupiter-mass planets orbit the host star, which are detected through direct imaging \citep{Marois+2008,Marois+2010}. From innermost to outermost, these planets are designated as Planet e, d, c, and b, with estimated deprojected separations of approximately 16, 27, 41, and 72 au, respectively, based on a combination of astrometric and dynamical models. In terms of mass, the innermost three planets are estimated to range from 7 to 10 $M_{\rm jup}$, with Planet d being the most massive, while the outermost planet has a mass of about 6 $M_{\rm jup}$ or less. However, the exact values depend on the model employed. Due to their substantial mass and close mutual separations, studies suggest that these planets are in a mean-motion resonance (MMR) chain of 8:4:2:1 to maintain system stability \citep{Fabrycky+2010, Esposito+2013, Konopacky+2016, Zurlo+2016, Wang+2018, Gozdziewski+2018, Gravity+2019, Gozdziewski+2020, Brandt+2021, Zurlo+2022, Thompson+2023}. The presence of planets in MMRs implies a potential history of convergent migration to reach their current positions \citep{Goldreich&Tremaine1980, Goldreich+2014}. Notably, only about 0.6\% of 0.1–3 $M_\odot$ stars host planets within the 5–13 $M_{\rm jup}$ range at orbital distances of 30–300 au \citep{Bowler2016}, making this architecture quite rare in the current context of exoplanets. The planets' atmospheres have also been characterized, revealing an unexpectedly high metallicity for all the planets \citep{Nasedkin+2024,Balmer+2025}.

Finally, the system hosts a debris disc, which was first detected by IRAS \citep{Sadakane2006} and initially spatially resolved using SPITZER \citep{Su+2009}. The disc comprises three components: an inner warm debris disc ($\lesssim$ 15 au), an outer cold debris disc ($\gtrsim 100$ au), and a mm-tail extending beyond the outer disc ($\gtrsim 300$ au), potentially indicating a scattered disc. The disc has been characterized across infrared, and (sub)-mm wavelengths \citep{Hughes+2011, Matthews+2014, Booth+2016, Faramaz+2021, Boccaletti+2024}.

All these features suggest a parallel with our Solar System, but on a larger scale: an inner asteroid belt analog, with four exterior giant planets, and a Kuiper belt analog beyond them. Due to these similarities with the Solar System, numerous models have been proposed to explain the formation and evolution of the HR~8799 system. Although many of these models have successfully explained certain aspects of the system, none have comprehensively accounted for all the observed features, as further discussed next.

\subsection{Discrepancies between observations and existing models} \label{sec:obs_mod_discrepancies}

The primary goals of existing models for HR 8799 are two-fold: (i) to explain the planetary MMR chain and (ii) to reproduce the structure of the debris disc, including the separation between the outermost planet's orbit and the inner edge of the outer debris disc. Although the first point is addressed by an important quantity of works (see Section \ref{sec:Architecture} for references), the second point remains elusive. 

Initial observations reported \rev{the outer disc's} inner edge \rev{to be} 100$\pm$10 au \citep{Su+2009,Matthews+2014}, consistent with the location of Planet b maintaining the edge, as predicted by the criterion of overlapping first order MMRs \citep{Wisdom1980}. Subsequent observations with higher angular resolution with ALMA at 1.3 mm \citep{Booth+2016} found larger radius for the inner edge, namely, $\approx 145$ au\footnote{This estimate, however, is debated. For instance \citet{Wilner+2018} report a value of $104_{-12}^{+8}$ au based on combined SMA and ALMA data.}. However, \citet{Wang+2018} calculated an inner edge of 93$^{+3}_{-2}$ au based on observational constraints related to Planet b's orbit. The inner edge at about 145 au was later corroborated by ALMA Band 7 observations \citep{Faramaz+2021}. Consequently, a new planet was proposed to explain and shape the disc, with $N$-body simulations suggesting a semi-major axis of approximately 100-120 au to maintain the edge \citep{Read+2018,Gozdziewski+2018,Pearce+2022}. Additionally, \citet{ImazBlanco+2023} found the inner edge to have a shallow surface-density profile, which is potentially consistent with collisional evolution rather than sculpting by a (hypothetical) non-migrating planet -- see also \citet{Pearce+2024}. However, this does not rule out sculpting by a migrating planet.

The dynamical stability of this hypothetical fifth planet was examined by \citet{Gozdziewski+2018}, who found that a planet with a mass of up to 3 $M_{\rm jup}$ could remain stable in an MMR with the other planets. However, observational constraints limit the maximum mass of such a planet to 1 $M_{\rm jup}$ \citep{Maire+2015,Zurlo+2022}. Alternatively, \citet{Read+2018} proposed reducing the mass of the fifth planet to as low as 0.04 $M_{\rm jup}$ \rev{(13 $M_\oplus$)} to achieve a strong alignment with observational data. Nevertheless, there is currently no evidence of this fifth planet in JWST/MIRI observations \citep{Boccaletti+2024}.


Finally, a notable characteristic of the disc is the possible presence of two distinct populations of bodies. \citet{Geiler+2019} used a collisional model to demonstrate that Herschel and ALMA data are incompatible without including an eccentric population. Similar to the Kuiper Belt, \rev{it seems that there is} a cold population, characterized by low eccentricities and low orbital inclinations (i.e., $e \sim 0.05$ and $ I < 5\degree$), and a hot or scattered population, composed of objects that would exhibit a wide range of eccentricities and/or orbital inclinations. A potential explanation for the hot population is provided by \citet{Gozdziewski+2018, Gozdziewski+2020}, who suggest that either a fifth planet is inducing resonant clumps or, alternatively, that the inward migration of Planet b excited bodies in the disc, which initially had an inner edge located at 150 au.

At this stage, it is evident that despite our extensive knowledge of the HR~8799 system and the numerous models put forward, these efforts are insufficient to fully explain its architecture—specifically, the cavity size, the shape of the inner edge, and the possible presence of scattered population. 
This highlights the need for reevaluating the mechanisms governing this system's formation and evolution. This work investigates and explores an alternative scenario to the typical convergent migration pathway, specifically focusing on how the super-Jupiters \rev{may have undergone outward migration, which} may shape the debris disc's radial profile while maintaining the planets' orbits in an interlinked MMR. We aim to enhance our understanding of the HR~8799 system by examining various configurations and migration scenarios.

\section{Methods}
\label{sec:methods}

\begin{table}
\centering

\begin{threeparttable}
\caption{Summary of parameters used to build up the fiducial simulation.}
\begin{tabular}{lcclcc}
\hline
\hline
\multicolumn{3}{c}{Gaseous disc } &
\multicolumn{3}{c}{Planetesimal disc }\\
\cmidrule(lr){1-3}
\cmidrule(lr){4-6}

Name & Unit & Value & Name & Unit & Value \\
\cmidrule(lr){1-3}
\cmidrule(lr){4-6}
$\Sigma_0$ & g cm$^{-2}$ & $27.17$ & $N$ & - & $10^4$ \\
$h_0$ & - & $0.05$ & $a_{\rm min}$ & au & 75 \\
$r_{\rm min}$ & au & 5 & $a_{\rm max}$ & au & 400 \\
$r_{\rm max}$ & au & 400 & $e$ & - & 0 -- 0.01 \\
$r_0$ & au & 10 & $I$ & degrees & 0 -- 0.60 \\
$P$ & - & 0.5 & $\omega$ & degrees & 0 -- 360 \\
$Q$ & - & 0.25 & $\Omega$ & degrees & 0 -- 360 \\
$T_d$ & Myr & 2.75 & $f$ & degrees & 0 -- 360 \\
$T_d^+$ & Myr & 0.90 & & &   \\

\hline
\multicolumn{6}{c}{}\\
\multicolumn{6}{c}{Planets}\\
\cmidrule(lr){1-6}
Name & Unit &Planet e &  \multicolumn{1}{c}{Planet d} & Planet c & Planet b \\
\cmidrule(lr){1-6}
$m_p$ & $M_{\rm jup}$ & 8.0 & \multicolumn{1}{c}{9.5} & 7.5 & 6.0 \\
$a_{\rm init}$& au & 15.12 &  \multicolumn{1}{c}{24.22} & 39.86 & 65.49 \\
$a_{\rm lim}$& au & 30.24 &  \multicolumn{1}{c}{48.44} & 79.72 & - \\
$e$& - & 0.163 &  \multicolumn{1}{c}{0.131} & 0.052 & 0.009 \\
$I$ & degrees & 0.0 &  \multicolumn{1}{c}{0.0} & 0.0 & 0.0\\
$\omega$ & degrees & 111.90 &  \multicolumn{1}{c}{202.33} & 129.24 & 305.81\\
$\Omega$ & degrees & 0.0 &  \multicolumn{1}{c}{0.0} & 0.0 & 0.0\\
$f$ & degrees & 233.45 &  \multicolumn{1}{c}{215.04} & 1.36 & 185.98\\
$\tau_{a}$ &Myr & - &  \multicolumn{1}{c}{-} & - & 8.0 \\
$\tau_{a}^+$ & Myr & 3.7 &  \multicolumn{1}{c}{2.7} & 1.3 & - \\
$\Tilde{\Gamma}$ & - & 0.75 &  \multicolumn{1}{c}{0.75} & 0.75 & - \\

\hline
\hline

\end{tabular}

\begin{tablenotes}
\small
\item \textbf{Gaseous disc.} It has a  surface density $\Sigma_0$ and aspect ratio $h_0$ at a reference radius of $r_{0}$, and   extends from $r_{\rm min}$ to $r_{\rm max}$. $P$ and $Q$ are the power-law indices for the density and aspect-ratio profiles, respectively (Equations \ref{eq:SD_profile_0} and \ref{eq:T_profile}). $T_d$ is the gas-disc-dispersal timescale, and $T_d^+$ is the outward-migration-acting timescale defined as a third of $T_d$.

\item \textbf{Planetesimal disc.} It is composed of $N$ massless particles with each particle initialized with a semi-major axis $a$ randomly such that they follow a power law $\propto a^{-0.5}$ from $a_{\rm min}$ to $a_{\rm max}$. The rest of the orbital elements, inclination ($I$), argument of periapsis ($\omega$), longitude of the ascending node ($\Omega$), and true anomaly ($f$) are randomly chosen. 

\item \textbf{Planets.} The planets have initial and fixed mass $m_p$ and initialized with a semi-major axis of $a_{\rm init}$. The inward migration timescale $\tau_a$ is defined for Planet b; meanwhile, the outward migration timescale $\tau_a^+$ is defined for the rest. These outward migrating planets have an upper limit for the semi-major axis $a_{\rm lim}$ and $\Tilde{\Gamma}$ is the normalized torque.

\end{tablenotes}
\label{tab:planets_params}
\end{threeparttable}
\end{table}





We conduct $N$-body simulations using {\sc Rebound} and its extension package, {\sc Reboundx} \citep{rebound,reboundx}, considering an HR 8799-like system consisting of a central star with mass $1.52\ M_\odot$, the four observed planets with masses from the inside out \rev{(}$\{8.0,9.5,7.5,6.0\}$ $M_{\rm jup}$\rev{)} and an outer debris disc composed of massless particles. The case with a potential fifth planet is presented in Section \ref{sec:five-planets scenario}.

We self-consistently consider the evolution of planets and a broad debris disc from the PPD phase, accounting for the gas's effects as a migration trigger on planets. We assume the planets are formed and initially in a resonant configuration, on orbits closer to the star than currently observed (see Table \ref{tab:planets_params}). The initial condition of the MMR chain enables the radial movement of all planets to be affected by changes in the movement of just one of them. We impose migration prescriptions for the planets along with a gas disc dissipation prescription, and we calculate the necessary migration timescales that the planets embedded in the PPD could have. 

\rev{Our aim is to explore a simple scenario, where we artificially impose migration and examine its effect on the planet and debris architecture. We emphasize that our model does not address the underlying causes or origins of migration. Rather, we aim to show that if some PPD mechanism(s) could drive outward and inward migration, then this could reproduce the observed system.}

Migration is a phenomenon that can be viewed as a competition between the local disc features and planetary mass; therefore, the amplitude and, importantly, the direction of migration will differ for different kinds of planets and discs. Consequently, we incorporate a prescription that can induce both inward and outward migration. A detailed description of the model, along with relevant justifications, is provided below. 


\subsection{\rev{Planet-m}igration prescription}
\label{sec:mig_pres_sub_sec}

Planetary migration \rev{can} arise from interactions between planets and a gaseous disc \citep{Lin&Papaloizou1979,Goldreich&Tremaine1979}. Disc torques facilitate the transfer of energy and angular momentum, typically leading to inward migration. While simulations  generally support this mechanism \citep[e.g.][]{Snellgrove+2001,Nelson&Papaloizou2002}, the process depends on disc parameters, such as viscosity and density. An important consequence of the migration process is the capture of the consecutive planets in MMR. 

$N$-body simulations can incorporate the migration effect as an imposed non-conservative force, as proposed by e.g. \citet{Beauge+2006}. 
For planet $i$, this force, per unit mass, can be written as:
\begin{equation}
    \boldsymbol{f}_i = \boldsymbol{f}_a\left(\tau_{a,i}\right) + \boldsymbol{f}_{e}\left(\tau_{e,i}\right),
\end{equation}
where the sub-indexes $i$ represent different planets in the system, 
and $\boldsymbol{f}_a\left(\tau_{a,i}\right)$ and $\boldsymbol{f}_{e}\left(\tau_{e,i}\right)$ the components related to the orbital decay and the circularization of the orbit, respectively. The parameters $\tau_{a,i}$ and $\tau_{e, i}$ are the radial and eccentricity damping timescales, respectively, which can be related via $\tau_{e,i}=\tau_{a,i}/\kappa$ \citep{Lee+2002}. We adopt $\kappa=100$ for our simulations.
By expanding the components of the force,
we obtain:
\begin{eqnarray}
    \boldsymbol{f}_a &=& -\frac{1}{2\tau_{a,i}}\boldsymbol{v}_i,\label{eq:f_a}\\
    \boldsymbol{f}_{e} &=& -\frac{2\kappa}{3\tau_{a,i}}\left[ \frac{1}{\left(1-e_i^2\right)}\boldsymbol{v}_i - \frac{\mu_i}{ \left|\boldsymbol{h_i}\right|}\left(\boldsymbol{\hat{h}}_i\times\boldsymbol{\hat{r}}_i\right)\right], \label{eq:f_e}\hfill
\end{eqnarray}
with $\boldsymbol{v}_i$ the velocity, $\boldsymbol{h}_i$ the specific angular momentum, $\boldsymbol{r}_i$ the position, $e_i$ the eccentricity, and $\mu_i = G(M+m_i)$ with $G$ the gravitational constant, $m_i$ the mass of the planet $i$, and $M$ is the mass enclos\rev{ed} by the planet.

Equations (\ref{eq:f_a}) and (\ref{eq:f_e}) \rev{would} lead to a continuous migration and eccentricity damping. To overcome this, we numerically implement an exponential growth for the migration timescale \citep{Gozdziewski+2018}, which is equivalent to an exponential decay in the force, such that:
\begin{equation}\label{eq:exponential_growth}
    \tau_{a,i}(t) = \tau_{a,i}(t=0)\ e^{t/T_d},
\end{equation}
where $T_d$ is the gas-disc-dispersal timescale or disc lifetime \citep{Mamajek2009}. Inherent to Equation \ref{eq:exponential_growth} is the assumption that the gas disc disperses uniformly over time. Then, different planets with different migration timescales will have the same exponential variation. The value of $T_d$ is observationally estimated to be between 1 and 10 Myr with a median value of 2--3 Myr \citep{Haisch+2001,Williams+2011}. We explore the values 1, 2.75, and 6 Myr to cover from short to \rev{long} disc lifetimes. 

Recent studies have challenged the traditional inward migration paradigm associated with classical migration theory. These findings suggest that outward migration may be a viable scenario under certain conditions. To model outward migration, we employ a modified version of Equations \ref{eq:f_a} and \ref{eq:f_e} \citep{Hahn+2005,Ali-Dib+2021}. To distinguish the parameters for outward migration from those of inward migration, we adopt a ``$+$'' superscript. Consequently, the forms of Equations \ref{eq:f_a} and \ref{eq:f_e} change to:
\begin{eqnarray}
    \boldsymbol{f}_a^+ &=& \left(\frac{1}{2\tau_{a,i}^+}\right)\left(\frac{a_{\rm lim}-a_{\rm init}}{a_i}\right)e^{-t/T_d^+} \boldsymbol{v}_i, \label{eq:f_a+}\\
    \boldsymbol{f}_e^+ &=& \boldsymbol{f}_e\left(\tau_{a,i}^+\right)\ e^{-t/T_d^+}      .
\end{eqnarray}
It is important to note that the change in sign in $\boldsymbol{f}_a$ will lead to exponential growth in the semi-major axis. To prevent this, the second term in Equation \ref{eq:f_a+} restricts the growth of the semi-major axis $a_i$ until it reaches a specific value, $a_{\rm lim}$, starting from an initial value, $a_{\rm init}$ (\rev{where ${a_{\rm init} < a_{\rm lim}}$}). However, this upper limit is unlikely to be attained due to the influence of the exponential term, which introduces exponential decay in the force. Additionally, we assume an eccentricity damping mechanism similar to that of the inward migration prescription, as described in Equation \ref{eq:f_e}, but also incorporating exponential decay of the force. The term $T_d^+$ defines the duration during which outward migration affects the planets, and we assume it to be a fraction of $T_d$. Consequently, we consider that outward migration concludes before inward migration ceases. This assumption is justified in Section \ref{sec:mech+migration}, where we discuss the  mechanisms that may trigger outward migration.



\subsection{\rev{Planet-m}igration regimes}
\label{sec:mech+migration}

The typical inward migration regimes are called type-I and type-II. Type-I migration affects low-mass planets that are unable to carve a gap around their orbits, while type-II migration applies to planets capable of opening a gap \citep{Papaloizou+2006,Nelson2018,Armitage2018}. Given the planetary masses in HR 8799, type-II migration is of interest. \citet{Kanagawa+2018} estimated the inward migration timescale as follows:
\begin{equation}\label{eq:tau_a}
    \frac{1}{\tau_{a}} = 6r_p^2\Omega_p\frac{q^2}{h ^2}\frac{\Sigma_{\rm gap}}{M_p},
\end{equation}
where the planet-related parameters $q=m_p / M_{\rm star}$, $r_p$, and $\Omega_p$ are the planet-to-star mass ratio, the planet's orbital radius, and angular velocity, respectively. Additionally, $h$ is the disc's aspect ratio, and $\Sigma_{\rm gap}$ is the gap's surface density given by:
\begin{equation}\label{eq:K}
    \Sigma_{\rm gap}=\frac{\Sigma_{\rm dd}}{1+\beta K},\quad\text{with}\quad K=\frac{q^2}{\alpha h^5},
\end{equation}
where $\Sigma_{\rm dd}$ represents the gas surface density in a disc-dominated case, in other words, it refers to the gas surface density that the disc would have in the absence of any perturbing planets, $K$ is the gap depth, $\alpha$ is the Shakura-Sunyaev viscosity \citep{Shakura&Sunyaev73}, and $\beta$ is a constant, which can be 0.034 or 0.040 according to \citet{Duffell+2013} or \citet{Kanagawa+2015}, respectively. We use $\beta = 0.034$ for our calculations.

On the other hand, outward migration can be reached in scenarios such as:

\begin{enumerate}[(i)]

    \item A Jupiter-like planet can migrate outward if the initial disc exhibits a steeply falling surface density profile \citep{Chen+2020},
    
    \item A planet can migrate if it is trapped at the edge of a gaseous disc when the inner rim of the disc moves outward due to photoevaporation \citep{Liu+2017,Liu+2022},

    \item A planet on an eccentric orbit can migrate outward when its eccentricity is $\geq 0.2$ \citep{DAngelo+2006},
    
    \item An eccentric gap ($\gtrsim0.1$) created by a super Jupiter-mass planet can drive the planet itself to migrate outward \citep{Dempsey+2021, Scardoni+2022}.
    
\end{enumerate}
The first two mechanisms encounter challenges when explaining the outward migration of planets more massive than Jupiter, as is the case in HR 8799. Furthermore, the second mechanism is only efficient for planets that are close to the star, making both unsuitable for application to the planets in HR~8799. In contrast, the third and fourth mechanisms are effective for supermassive planets and do not face these limitations. However, an initially high-eccentricity planet is rapidly circularized due to the influence of the PPD \citep{Bitsch2013}, making the third scenario unlikely. \rev{Although the fourth mechanism seems promising, it is important to note that all the mechanisms discussed so far consider the outward migration of a single planet. Introducing additional planets could alter the net torque on the system, potentially changing its magnitude or even the direction of migration. They could also affect the gap profiles, or merge gaps together. Nonetheless, we use the fourth mechanism as a reasonable basis for our hypothetical migration prescription, noting again that we are less concerned with the physical origins of migration and more with its potential impact on the system architecture.}

According to the fourth mechanism, when a single planet possesses sufficient mass ($\geq 3\ M_{\rm jup}$), it can carve out an eccentric cavity, which increases its accretion rate and generates a positive torque over the planet \citep{Kley+2006,Fung+2014,Tanaka+2022}. This torque induces outward planetary migration, with an associated timescale given by:
\begin{equation}\label{eq:tau_a+}
    \frac{1}{\tau_{a}^+} =12\pi\ \Tilde{\Gamma} \alpha r_p^2\Omega_p h^2\frac{\Sigma_{\rm dd}}{M_p}, 
\end{equation}
where $\Tilde{\Gamma}$ is the normalized torque such that $\Tilde{\Gamma}\in(0,1]$ according to numerical simulations \citep{Dempsey+2021,Scardoni+2022} \rev{to ensure an outward migration. This value may change with the addition of extra planets in their calculations.}. The increase in the accretion rate, approximately an order of magnitude greater than that of a non-eccentric gap, would lead to rapid gas depletion in the planet's vicinity. If multiple planets in adjacent orbits are considered under this condition, a significant expansion of the inner PPD cavity is plausible. However, this rapid accretion phase would likely halt outward migration due to either the lack of gas material to exert torque on the planet or the circularization of the gap. This sequence of events would support the assumption that the outward migration process is relatively short-lived compared to the life span of the gas disc, thereby our choice of $T_d^+ \lesssim T_d$ (Section \ref{sec:mig_pres_sub_sec}).

In Figure \ref{fig:K_alpha}, we compute the transition inward/outward migration limit in terms of $K$ (see Equation \ref{eq:K}) as a function of the $\alpha$-viscosity to determine the planetary migration regime for the planets in HR~8799. The figure shows that a range of \rev{$\alpha \approx \left[3{\small-}8\right] \times 10^{-4}$} would drive outward migration of the inner planets e, d, and c, while Planet b experiences inward migration.

\subsection{Simulation setup}

\begin{figure}
\centering
\begin{center}
    \includegraphics[width=0.48\textwidth]{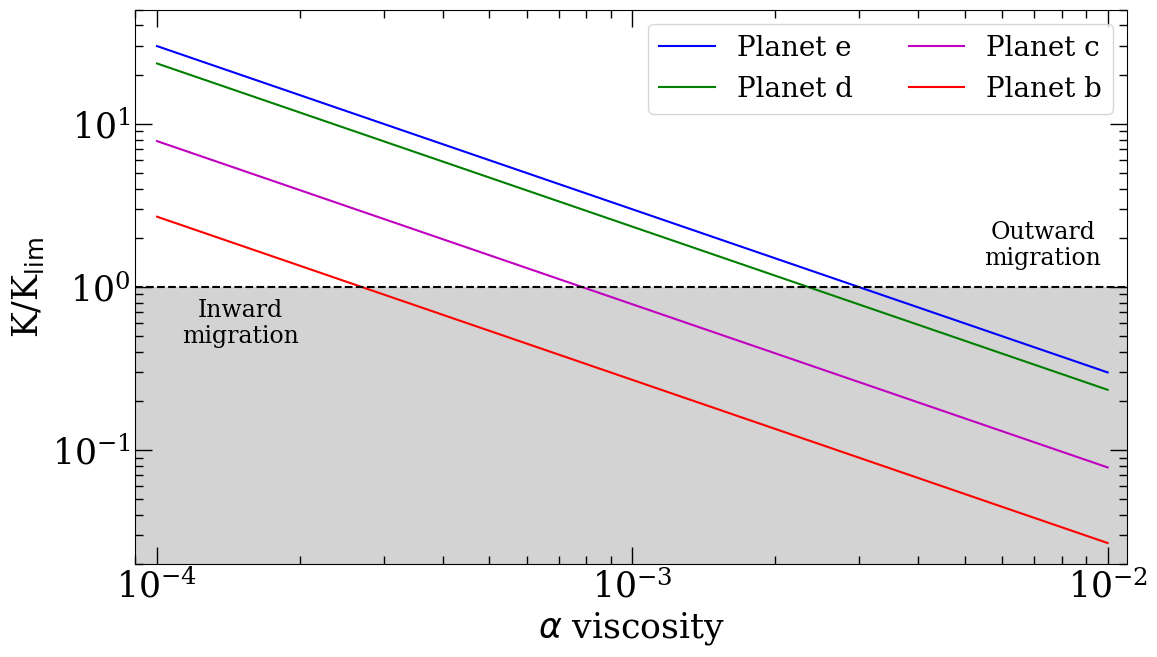} 
    \caption{Dependence of the $K$ (see Equation \ref{eq:K}) value for all four planets in HR~8799, which sets that planet's migration direction as a function of $\alpha$-viscosity. Values are shown relative to $K_{\rm lim}=1.5\times10^4$ \citep{Scardoni+2022}, which marks the transition between a planet migrating inward or outward, highlighted as gray and white regions, respectively. 
    }
    \label{fig:K_alpha}
\end{center}
\end{figure} 

\begin{figure*}
\centering
\begin{center}
    \includegraphics[width=1.0\textwidth]{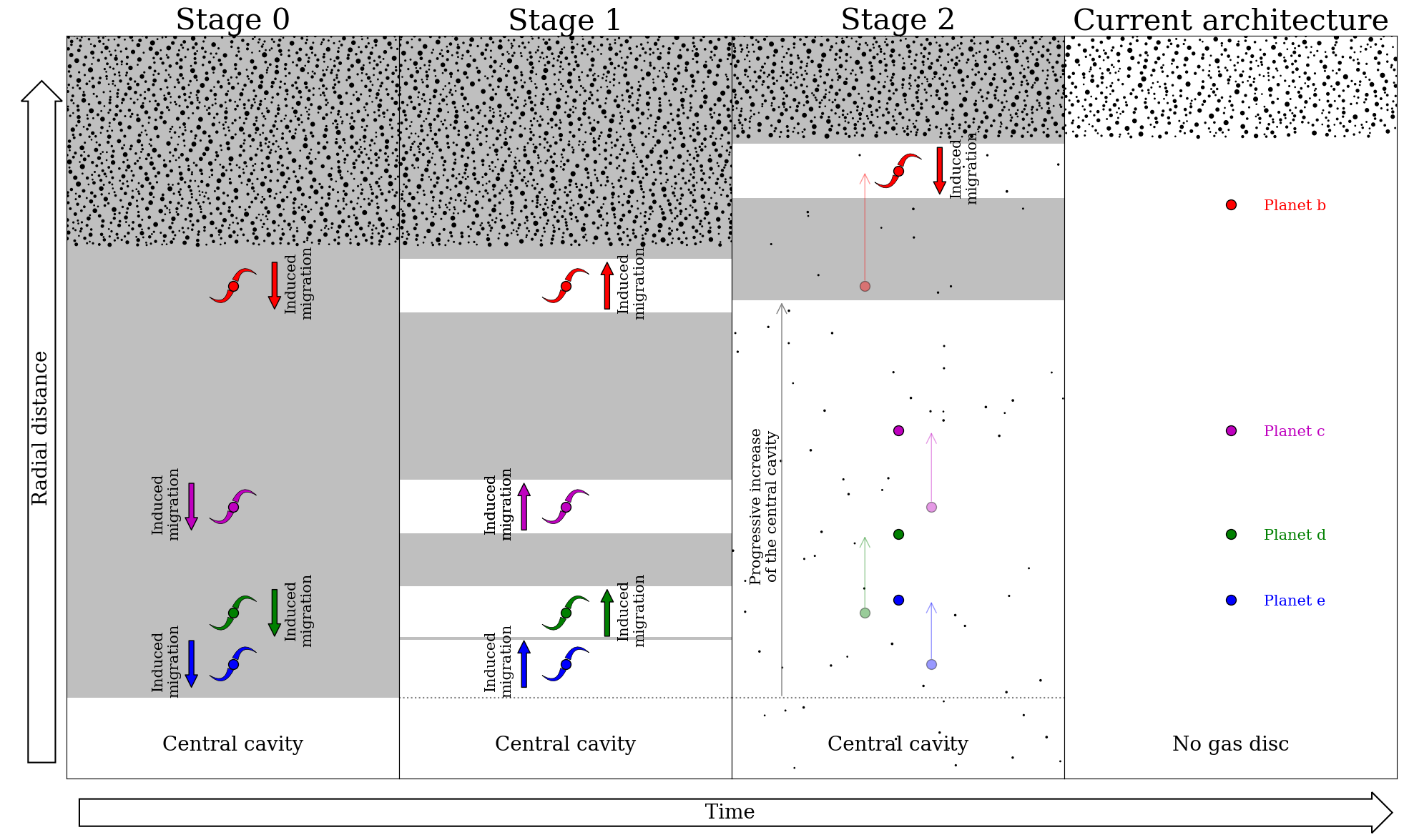} 
    \caption{
    Schematic representation of the proposed evolutionary model for the HR~8799 system. The pair of curves around planets represents the planet under the influence of the primordial gas disc, which is represented as the gray region. By contrast, the gas-free region is colored in white. Finally, the black dots represent the planetesimal bodies. In Stage 0, the protoplanets form and migrate inward until they are stopped at the inner edge and trapped in an MMR. In Stage 1, the three inner planets migrate outward, which compels the outermost planet to also migrate outward due to their MMR chain, even though it would otherwise move inward due to gas drag. This outward movement \rev{displaces} the inner edge of the planetesimal disc. In Stage 2, the three inner planets cease their outward migration, allowing the outermost planet to resume migrating inward, although this movement is marginal. At this point, the system achieves its current architecture.
    }
    \label{fig:sketch}
\end{center}
\end{figure*} 

The protoplanetary disc is assumed to have an initial surface density $(\Sigma_{\rm dd})$ and aspect ratio $(h)$ profiles as follows:
\begin{eqnarray}
    \Sigma_{\rm dd}(r) &=& \Sigma_0\ \left(\frac{r}{r_0}\right)^{-P}\left(1-\sqrt{\frac{r_{\rm min}}{r}}\right), \label{eq:SD_profile_0}\\
    h(r) &=& h_0\ \left(\frac{r}{r_0}\right)^{Q} \label{eq:T_profile},
\end{eqnarray}
where $r_0$ is a reference radius, while $\Sigma_0$ and $h_0$ are the surface density and aspect ratio values at $r_0$, respectively. The disc parameters, planets, and planetesimal disc are in Table \ref{tab:planets_params}. Based on Section \ref{sec:mech+migration} and Figure \ref{fig:K_alpha}, we set $\alpha=5\times10^{-4}$ to have an outward migration of the three interior planets and an inward migration of the outermost planet. We build the initial condition by \rev{setting} the planetary orbital parameters to be in MMR with period ratios 8:4:2:1, assuming prior short-term migration had driven the system into this resonant state. 
The feasibility of this assumption is discussed in Section \ref{sec:discussion}. 

\rev{We do not impose gas drag on planetesimals, because it would have a negligible effect compared to other forces. To demonstrate this, we use \mbox{Equation 12} in \cite{Best+2024} (based on \citealt{Adachi+1976}) to calculate the expected planetesimal-migration timescale due to gas\footnote{\rev{Note that $\rho$ in \mbox{Equation 12} of \cite{Best+2024} actually denotes gas density, rather than planetesimal bulk density as written in that paper.}}. We use the system parameters in \mbox{Table \ref{tab:planets_params}}, and assume the drag coefficient ${C_{\rm D} \sim 1}$. Using this, we calculate that a ${1 \; \rm km}$ planetesimal at ${150 \; \rm au}$, with a bulk density of ${1 \; \rm g \; cm^{-3}}$, would have a migration timescale of ${\sim \rm Gyr}$. The timescale is even longer for larger bodies. These are far longer than the expected ${<10 \; \rm Myr}$ gas-disc lifetime, so we conclude that gas drag would have a negligible effect on planetesimals in our scenarios.}


We use {\sc Mercurius} \citep{Rein+2019}, a hybrid integrator that switches between the symplectic {\sc WHFast} when particles are well separated to {\sc IAS15} when a close encounter occurs. The timestep is set to 0.1 yrs but can decrease down to $10^{-4}$ yrs when a close encounter is detected. We evolve the simulation until it reaches 20 Myr. 



\section{Results}
\label{sec:results}

This section is divided into five parts: in Section \ref{sec:overview} we present an overview of the proposed evolutionary pathway of HR~8799. Section \ref{sec:parameter_dependencies} explores the general applicability of the model concerning different sets of parameters. Section \ref{sec:fiducial_model} examines the fiducial model for four planets, showcasing outcomes from a simulation constructed using the parameters outlined in Table \ref{tab:planets_params} while Section \ref{sec:five-planets scenario} \rev{extends} the fiducial model by incorporating a fifth planet. Finally, in Section \ref{sec:sim_obs}, we compare the simulated planetesimal radial profile with the ALMA Band 7 observation \citep{Faramaz+2021}.

\subsection{Overview of the evolutionary phases in the HR 8799 system}\label{sec:overview}

Figure \ref{fig:sketch} provides a broad overview of our proposed multi-stage evolution model for the planets in HR~8799, highlighting their correlations with the physical processes occurring in the disc, which we unpack below:

\begin{itemize}
    \item \textit{Stage 0}: We assume classical convergent migration of the protoplanets, similar to the formation mechanism explored by \citet{Gozdziewski+2018}. However, in our model, the four planets are trapped in resonance at smaller distances than those currently observed; the semi-major axis value for Planet b is about 10\% smaller than the current value. These configurations serve as our initial conditions for running the simulations, allowing the outermost planet to push outward the inner edge of the planetesimal disc when it begins to migrate in Stage 1.
    
    \item \textit{Stage 1}: The three inner planets migrate outward as the gas disc gradually dissipates. The inner planets will continue migrating until there is no gas to exert torque. As discussed in the Section \ref{sec:mech+migration}, the gas depletion is numerically simulated by incorporating the terms $T_d$ and $T^+_d$ with  $T_d>T^+_d$. Note that we are indirectly modeling a faster gas depletion \rev{for} the three innermost planets with this assumption. The outward movement of these planets also compels Planet b to migrate outward due to MMR, and despite its inward migration prescription, because of the gas. Additionally, the \rev{inner edge of the} planetesimal belt is \rev{displaced as} Planet b \rev{moves out into it}, resulting in the generation of high-eccentricity bodies\rev{, some of which} will subsequently enter the disc's inner regions, i.e., regions spanning between the star and Planet b.
    
    \item \textit{Stage 2}: The new architecture is \rev{now} wider than the initial configuration but not in alignment with the observed structure. The inner gas disc has nearly \rev{been depleted}, except in the \rev{outer} regions close to Planet b\rev{, similar to a cavity in a transition disc}. This remaining gas will drive Planet b to migrate inward. Following this final migration, the architecture aligns with the current locations of the planets. All planets were likely subjected to migration phases characterized by high accretion rates to achieve the final observed state. For simplicity and due to model limitations, we maintain constant planetary masses and assume a uniform gas disc profile. However, these simplifications could affect the planet-disc evolution, a topic further explored in Section \ref{sec:discussion}.

\end{itemize}

In the following Section, we will detail the behavior of all aspects and this fiducial model evolution, including a five-planet model.

\begin{figure}
\centering
\begin{center}
    \includegraphics[width=0.48\textwidth]{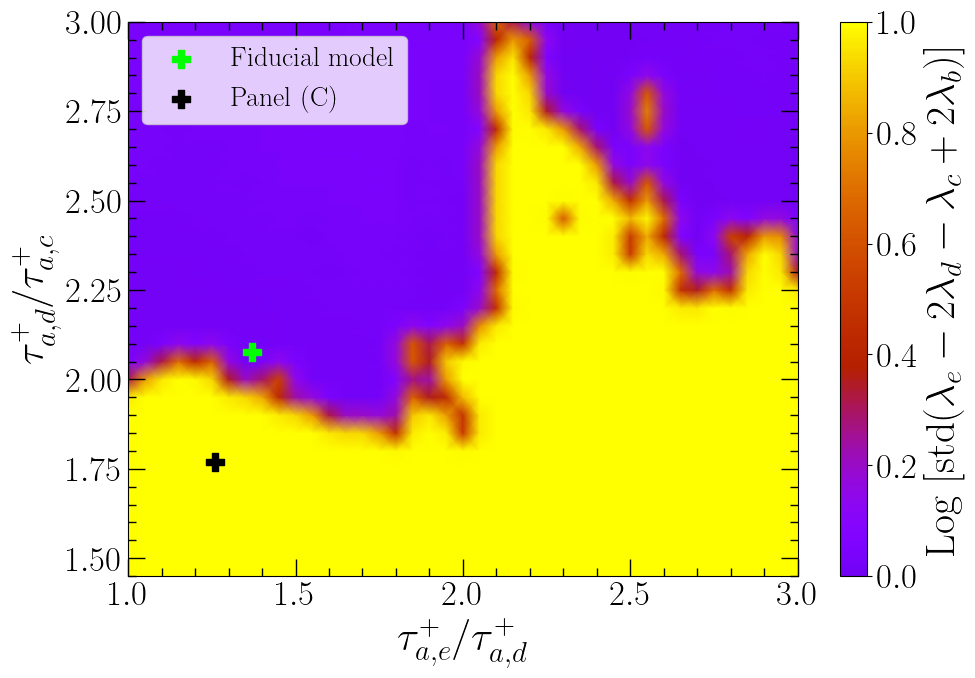} 
    \caption{Dependence of maintaining the MMR chain among the three innermost super-Jupiter planets as a function of the ratios $\tau_{a,e}^+/\tau_{a,d}^+$ and $\tau_{a,d}^+/\tau_{a,c}^+$. The plot exhibits a grid composed of $20\times40$ simulations. The color scale represents the standard deviation normalized to the minimum value of the resonant angle $\lambda_e - 2\lambda_d - \lambda_c + 2\lambda_b$ over 20 Myr of evolution. Thus, values close to 0 indicate that the MMR is preserved throughout the entire evolutionary period. Here, $\tau_{a,c}^+$ and $T_d$ are fixed at 1 and 2.75 Myr, respectively, while $\tau_{a,e}^+$ and $\tau_{a,d}^+$ are varied. The green and black crosses represent the values that correspond to the fiducial model and the Panel (C) simulation in Figure \ref{fig:T_ds}.}
    \label{fig:Tau_dependence}
\end{center}
\end{figure}

\subsection{Parameter dependencies and model constraints}
\label{sec:parameter_dependencies}

\begin{figure*}
\centering
    \includegraphics[width=0.8\textwidth]{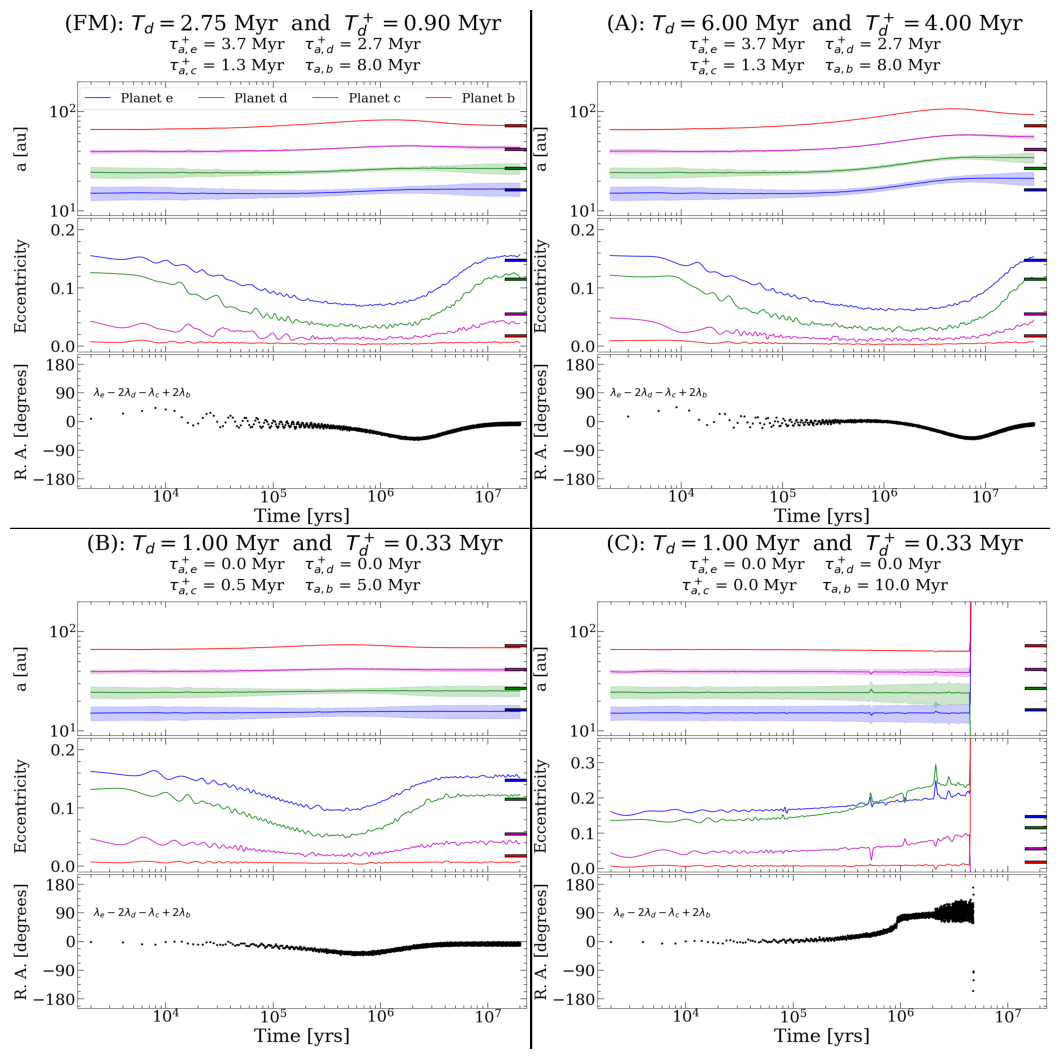} 
    \caption{Four examples illustrating the dependence among the final planetary state, the disc lifetime ($T_d$), and the outward migration duration ($T_d^+$) along with the different values of $\tau_{a,i}$ and $\tau_{a,i}^+$. In each case, the upper panel shows the evolution of the semi-major axis (a) with shadows representing the variation in the pericentre and apocentre distances, the middle panel displays the eccentricity, and the bottom panel illustrates the evolution of the resonant angle (R.A.) for the four-planet architecture. The color-code used is the same as in Figure \ref{fig:sketch}. The estimated semi-major axes and eccentricities \rev{of the observed planets} are indicated by horizontal lines on the right border, which match the corresponding planet colors. \textbf{Panel~(FM)} represents the fiducial model for four planets with the parameters listed in Table \ref{tab:planets_params}. \textbf{Panel~(A)} presents a scenario \rev{with a longer disc lifetime and outward-migration duration,} with $T_d = 6.00$ Myr and $T_d^+ = 4.00$ Myr until a final time of 30 Myr\rev{. I}t considers the same values for $\tau_{a,i}$ and $\tau_{a,i}^+$ as in the fiducial model. \textbf{Panel~(B)} displays a model where the migration timescales for the two innermost planets are set to 0 Myr\rev{, whilst} the migration evolution is maintained for the two outermost planets, \rev{as could arise if an inner cavity were present. T}he values $T_d$ and $T_d^+$ \rev{are} set to 1.00 and 0.33 Myr, respectively. \textbf{Panel~(C)} illustrates a model where the three innermost planets do not migrate, but only the outermost does inward.}
    \label{fig:T_ds}
\end{figure*} 

Our model relies on the parametrization of the primordial gaseous disc to impose migration regimes on the planets and planetesimals. Naturally, this approximation strongly depends on the parameters used to model the physical interactions. This section explores the impact of varying the most relevant parameters on the final outcome. The model encapsulates planet-disc interactions in the quantities $\tau_{a,i}$, $\tau_{a,i}^+$, and $\tau_{e,i}$, which are crucial for modeling migration. Additionally, the dispersion of the gas disc will determine the migration duration, $T_d$, as well as the duration of the outward migration, $T_d^+$. 


We examine how the MMR chain responds to variations in the outward migration timescale of the three inner planets. Figure \ref{fig:Tau_dependence} shows the result of a grid $20\times40$ different initial conditions integrated over 20 Myrs in the ($\tau_{a,e}^+/\tau_{a,d}^+$, $\tau_{a,d}^+/\tau_{a,b}^+$) plane. This figure is not a stability map akin to MEGNO \citep{Cincotta&Simo2000}; instead, it depicts a measure of the dispersion of the resonant angle $\lambda_e - 2\lambda_d - \lambda_c + 2\lambda_b$ computed as the standard deviation for different simulations, where $\lambda_{i}$ is the mean longitude of a planet $i$. Simulations exhibiting a large dispersion, i.e., values close to 1, do not belong to a four-planet MMR but may still maintain stable orbits. The crosses represent simulations built in these regimes: the fiducial model in the stable MMR region in green and a simulation inside the unstable zone in black.

As detailed in the Section \ref{sec:methods}, we have selected a set of reasonable parameters for both planets and discs that yield a stable MMR chain. However, this is not a unique or exotic result. Our setup employs parameters within a notorious resonant-keeping region, violet regions in Figure \ref{fig:Tau_dependence}. The broad region that keeps the MMR suggests that different initial disc conditions could promote both outward and inward migration behavior, similar to what we look for in the fiducial model. While initial conditions of the PPDs remain a topic of ongoing debate, it can be argued that a chain of four super-Jupiter planets could maintain resonances following an outward migration phase under typical initial disc conditions.


A favorable outcome for the MMR chain may not exactly reproduce the final values of the planets' semi-major axes and eccentricities. The variables $T_d$ and $T_d^+$ have an important impact on the final outcome. Some examples of varying these parameters are displayed in Figure \ref{fig:T_ds}, which we unpack below.
\begin{itemize}

    

    \item Panel (FM). This panel represents the evolution of the fiducial model using the parameters outlined in Table \ref{tab:planets_params}. It illustrates the archetypal evolution depicted in Figure \ref{fig:sketch} and will be explained in detail in Section \ref{sec:fiducial_model}. 

    \item Panel (A). This panel presents the four-planet evolution for a scenario with a longer disc lifetime and outward migration duration. In the regime where $T_d = 1.5 T_d^+$, there is a prominent increase in the semi-major axis values, with Planet b experiencing roughly a 62\% increase from its initial value to its peak due to the extended duration of the outward migration mechanism. However, this scenario requires more time to achieve a steady state without planetary migration movements. In this scenario, the final eccentricity values align with those observed; however, the semi-major axes exceed the expected values.


    \item Panel (B). This panel illustrates the evolution of the four planets when the two innermost planets are neither set to migrate inward nor outward, i.e., $\tau_{a,i} = \tau_{a,i}^+ = 0$ Myr, while Planet c is set to migrate outward ($\tau_{a,c}^+ = 0.5$ Myr) and Planet b inward ($\tau_{a,b} = 5$ Myr). This scenario \rev{could be physically} achieved by introducing a larger inner cavity encompassing both innermost planets while the other two remain embedded in the gas disc. To compute the mentioned migration timescales of the migrating planets consistently, we use the same gaseous disc parameters as in Table \ref{tab:planets_params}, but adjusting $P = 1$, $\Tilde{\Gamma} = 1$, and $T_d = 1$ Myr while maintaining the same ratio for $T_d^+$ than the fiducial model, resulting in a value of 0.33 Myr. The outcomes are similar to what we look for the fiducial model with two differences: the increase from the initial value of Planet b's semi-major axis to its peak is about 12\%, half of that of the fiducial model, and the time to reach a steady state is significantly short, achieving it by almost 4 Myr.  

    \item Panel (C). This panel illustrates the evolution of a four-planet system where the three innermost planets are not set to migrate either inward or outward, while Planet b is set to migrate inward with $\tau_{a,b} = 10$ Myr. This scenario represents the possibility of having the three innermost planets located inside the central gas cavity, while the outermost planets remain embedded in the gas disc. In this setup, the planets gradually acquire eccentricities and slowly migrate inward until, by 5 Myr, the entire system is disrupted.
    
\end{itemize}

Changes in $T_d$ and $T_d^+$ can lead to varying outcomes for the same set of $\tau_{a,i}$ and $\tau_{a,i}^+$ values. These initial conditions related to both the planets and the disc can be freely explored, with the constraint of achieving the final state observed in HR~8799. The fiducial model shows that an alternative evolutionary pathway for the system is feasible. However, it's important to note that the outcomes derived from these models are not unique. Similar migration behavior to the fiducial model can also be observed in other scenarios, such as cases (A) and (B), indicating that various configurations and pathways, though not exotic, might lead to comparable and stable evolutionary results. Nevertheless, the planetary system may become unstable and disrupted if the three innermost planets reside in the cavity and only the outermost planet migrates inward, as illustrated in case (C).

The case (A) displays a good match in the eccentricity values but not in the semi-major axes, which are larger than those expected. This discrepancy could be resolved by making the initial planetary architecture even more compact, allowing the expected values to be reached during planetary evolution. On the other hand, the fiducial model assumes that each planet has its own gap, which may be difficult to sustain for a pair of neighboring massive planets that likely would share a common gap \citep{Dong+2015,Kanagawa+2018,Kanagawa+2020}. This scenario is illustrated in case (B), where the two innermost planets are thought to be in the central cavity without interactions with the gas disc. Therefore, only planets c and b will experience outward and inward migration movements respectively.

\begin{figure}[ht]
\centering
    \includegraphics[width=0.48\textwidth]{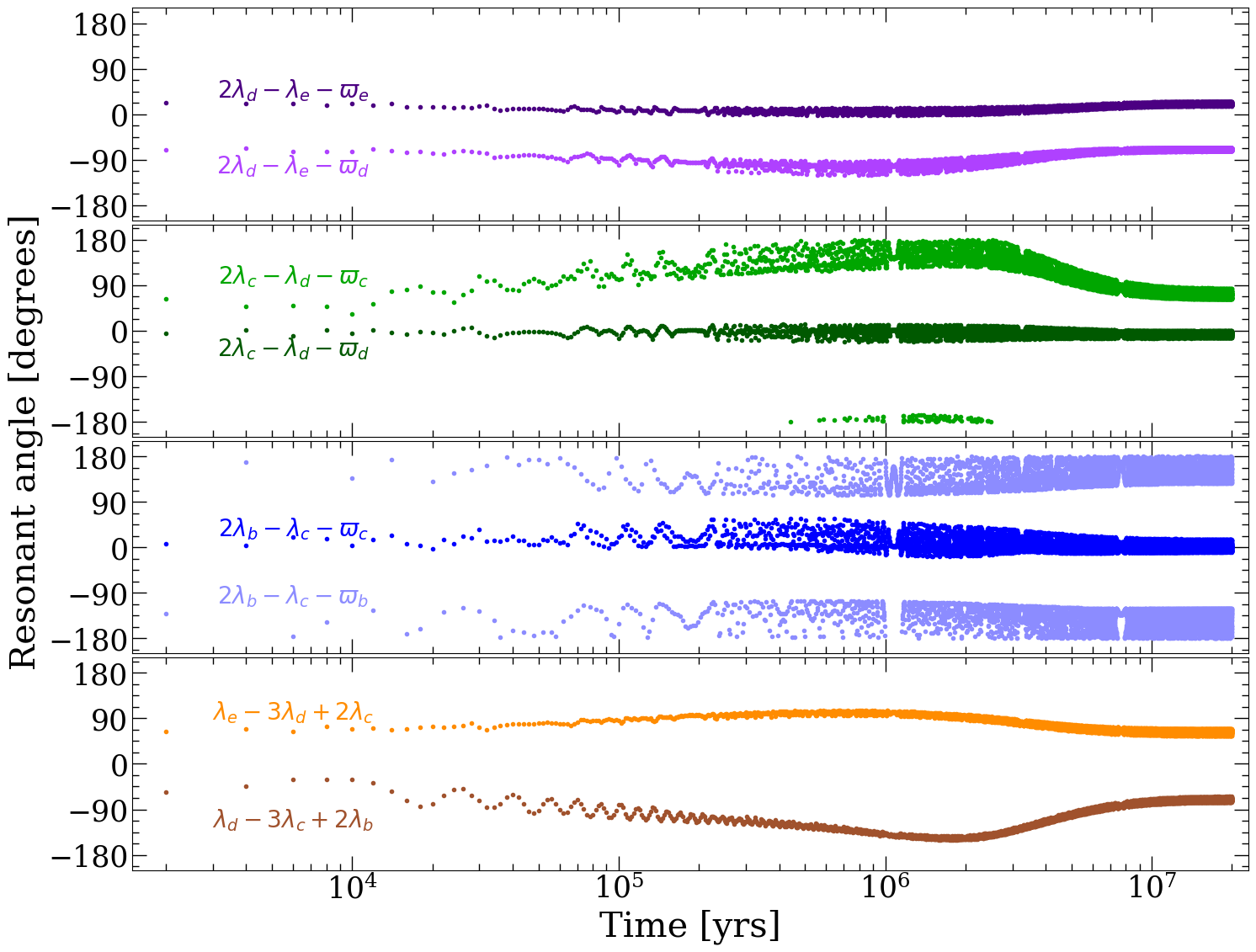}\\
    \includegraphics[width=0.48\textwidth]{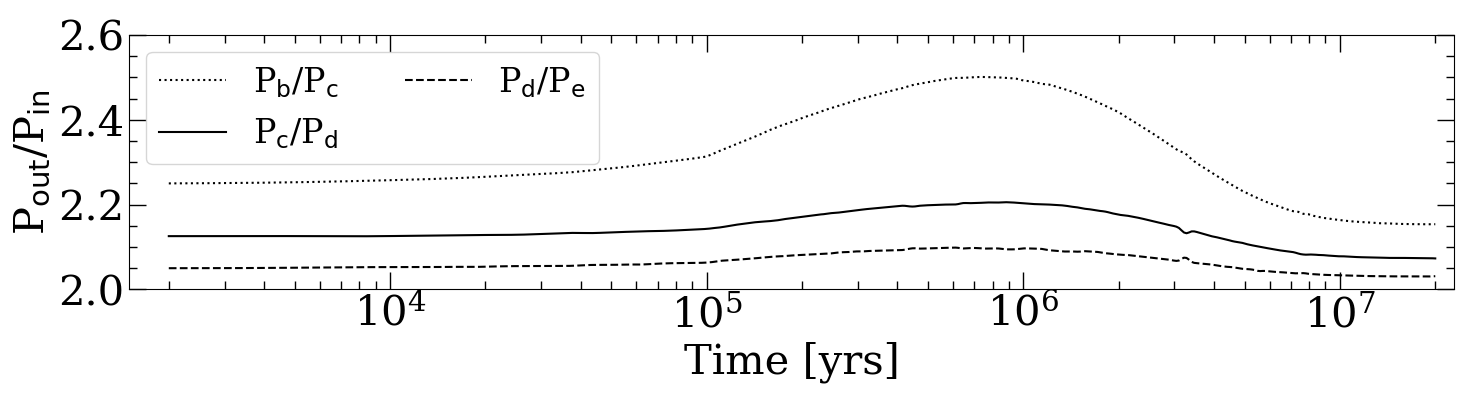} 
    \caption{Resonant angles and period ratios evolution for adjacent planets \rev{in the fiducial model}. \textbf{Upper panels}: The resonant angle for each adjacent pair and trio of planets is shown. \textbf{Bottom panel}: The evolution of the period ratio of two adjacent planets. }
    \label{fig:main_plot}
\end{figure} 


\begin{figure*}
\centering
    \includegraphics[width=1.0\textwidth]{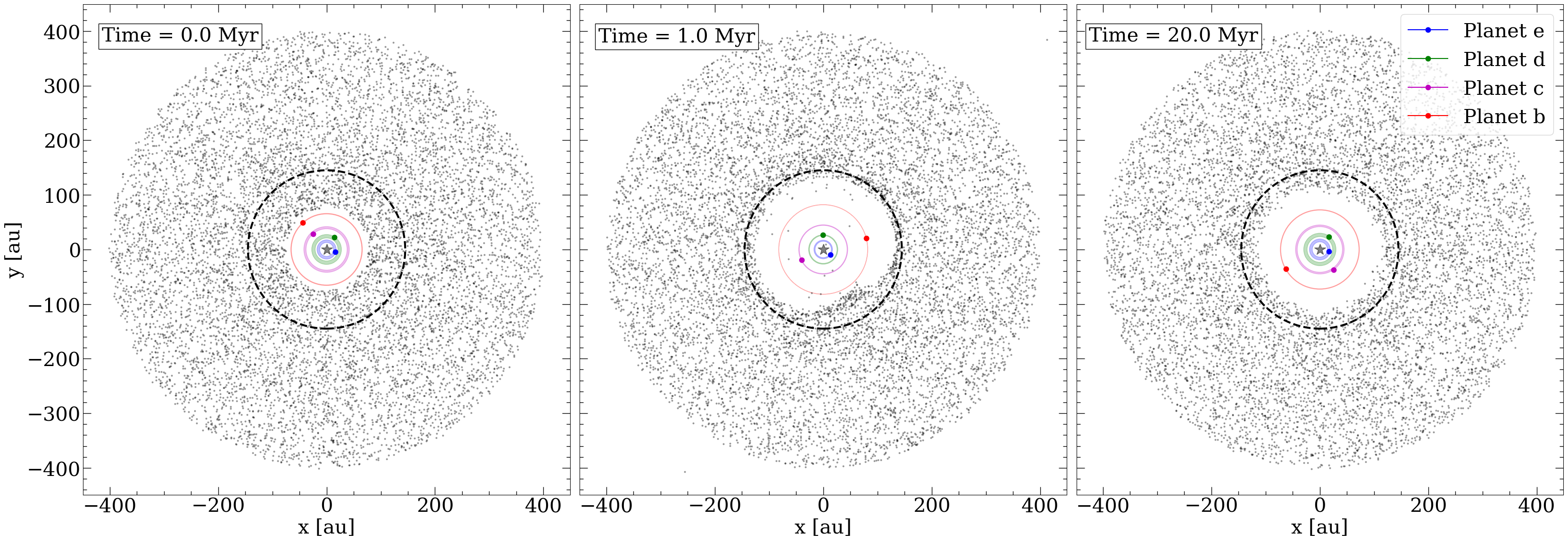}\\ \includegraphics[width=1.0\textwidth]{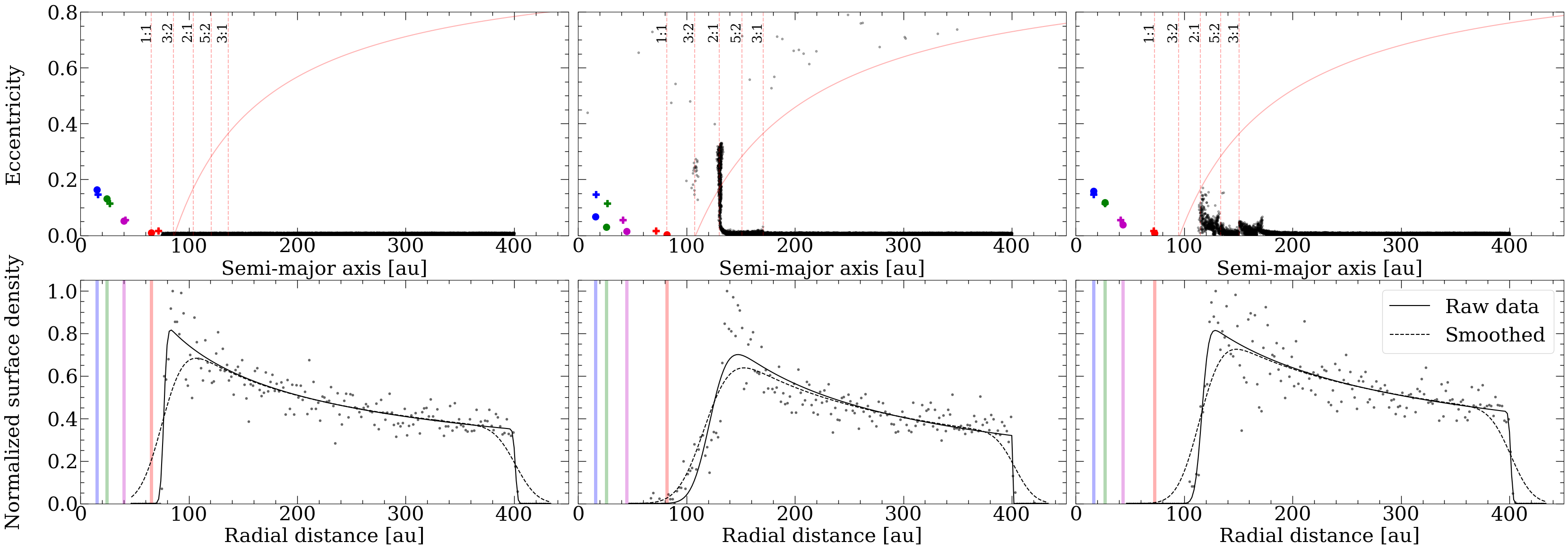} \\
    \caption{Snapshots showing the evolution of planetesimals at 0, 1.0, and 20 Myr from left to right, respectively. \textbf{The upper row} displays the planetesimal distribution in the xy-plane, the planets in colors (color-coded as in Figure \ref{fig:sketch}), and the dashed black circle represents the \rev{proposed 145 au value of the observed} central cavity. \textbf{The middle row} shows the $ae$-plane as the colored circles with the instantaneous values and the crosses with the observed values. The vertical lines represent the location of MMR for the 1:1, 3:2, 2:1, 5:2, and 3:1 resonances referring to the outermost planet (Planet b), and the red curve represents the minimum eccentricities required for debris to come within 3 Hill radius of the Planet b.
    \textbf{The lower row} shows the normalized surface density profile. The dots indicate the radial bins for surface density, while the black-solid and black-dashed lines represent the fitted curve according to Equation \ref{eq:SDprofile} for the raw data and the smoothed by a Gaussian beam size of 43 au, respectively. The locations of the planets are marked by vertical lines for reference.}
    \label{fig:main_plot_2}
\end{figure*}

\subsection{Fiducial model - four planets}
\label{sec:fiducial_model}

As mentioned, Figure \ref{fig:T_ds} (panel FM) shows the evolution of the four planets' semi-major axes, eccentricities, and the four-planet resonant angle.  Complementarily, Figure \ref{fig:main_plot} displays the evolution of resonant angles for pairs and triplets of adjacent planets, along with their period ratios. Finally, Figure \ref{fig:main_plot_2} depicts the planetesimal disc's state at three different times. 

\subsubsection{Four-planets architecture evolution}

All planets experience outward movement when outward migration begins, even though Planet b is configured to migrate inward. This phenomenon can be explained by comparing the migration timescales of Planets b and c, with $\tau_{a,b}/\tau_{a,c}^+ \approx 8$. 
This indicates that the outward movement of Planet c is faster than Planet b's inward migration. During the first $\sim$ 1Myr, the eccentricities of the three innermost planets decrease, while the Planet b's value remains relatively unchanged. For this time, the period ratios of all consecutive pairs increase. When the outward migration stops, Planet b begins to migrate inward, the moment at which the migration turns convergent, and the period ratios begin to decrease.  

A crucial aspect to note is that even during short-term divergent migration, the resonant chain is preserved, making the system migrate as a block. Although the orbital period ratio for the outermost planetary pair increases to about 2.5, the resonant angle in all configurations continues to oscillate around a fixed value. \textcolor{black}{This deviation from exact commensurability is not inconsistent with resonance; in fact, such behavior is expected at low eccentricities, where the width of the resonance in period ratio increases. Moreover, the center of libration of the resonant angle can shift significantly away from the nominal 2:1 value, depending on the masses and mass ratios of the planets \citep[see, for example, ][]{2022MNRAS.514.3844C}.}

At the end of the outward migration, Planet b begins to migrate inward due to the imposed condition $T_d > T_d^+$. The peak in the outward displacement is reached at 25\% of its initial position and continuously decreases later. This inward migration induces the other planets to increase their eccentricities. This effect, referred to as the pantographic effect in \citet{Cerioni&Beauge2023}, is similar to a train, where the entire train moves due to the action of a single locomotive. The ability of Planet e to be influenced by Planet b is facilitated by the resonance chain among the planets; without this resonance, the effect would occur sequentially, progressing inward in pairs from the outermost regions. The inward movement results in a compact planetary architecture as the beginning: the planets experience a modest reduction in their semi-major axes, their eccentricities increase because of the resonant capture, and they approach their initial period ratios relative to their neighbors.

\begin{table}
\centering
\begin{tabular}{c|cccccc}

\hline

\hline
\hline
\multicolumn{5}{c}{}\\
\multicolumn{5}{c}{Four planets}\\
\cmidrule(lr){1-5}

Time & $r_{\rm in}$ & $\sigma_{\rm in}$ & $\tilde{r}_{\rm in}$ & $\tilde{\sigma}_{\rm in}$ \\
Myr & au & - & au & - \\
\hline
0.0  & 77.01 & 0.03 & 77.69 & 0.20\\
1.0  & 122.99 & 0.10 & 119.65 & 0.16\\
20.0 & 116.44 & 0.04 & 116.05 & 0.14\\
\hline 
Observed & 160 & $0.27^{+0.04}_{-0.08}$ & 160 & $0.27^{+0.04}_{-0.08}$\\
\hline
\multicolumn{5}{c}{}\\
\multicolumn{5}{c}{Five planets}\\
\cmidrule(lr){1-5}
Time & $r_{\rm in}$ & $\sigma_{\rm in}$ & $\tilde{r}_{\rm in}$ & $\tilde{\sigma}_{\rm in}$ \\
Myr & au & - & au & - \\
\hline
0.0  & 112.96 & 0.00 & 113.00 & 0.15\\
1.0  & 133.06 & 0.15 & 133.60 & 0.18\\
20.0 & 153.84 & 0.10 & 152.19 & 0.14\\
\hline 
Observed & 160 & $0.27^{+0.04}_{-0.08}$ & 160 & $0.27^{+0.04}_{-0.08}$\\
\hline
\hline

\end{tabular}
\caption{The surface density fitted parameters for the disc's inner edge are provided for three different evolutionary times, considering both four- and five-planets, along with the observationally measured parameters for the debris disc in the HR 8799 system \citep[see table 1 of][]{Pearce+2024}. The tilde-marked parameters are derived from fitting over a smoothed density profile.}
\label{tab:4planetsSD}
\end{table}

\subsubsection{Planetesimal belt morphology}

The planetary evolution effects on the planetesimal disc are significant, particularly at the inner edge, as shown in Figure \ref{fig:main_plot_2}. The initial outward movement of Planet b \rev{displaces} planetesimals \rev{initially} located between 75 au (the inner rim) and the position where the outer 2:1 resonance takes place at the beginning ($\sim 104$ au). \rev{These planetesimals are either ejected through scattering or swept into MMRs, where they may increase in eccentricity until they cross the planets' orbits}. This interaction could result in either ejection from the system or collisions with the planets \citep{Wyatt+2017}, although collisions are not modeled in this work. Such collisions, triggered by orbital instabilities, would be similar to those proposed in the Nice model, contributing to the Late Heavy Bombardment \citep{Bottke+2017}. The crossing-orbit population can be observed at time 1.0 Myr, though the regions near the planets are nearly cleared by the end of the simulation. 
Notably, one of the most affected planetesimals populations is the one trapped in the 3:2 resonance, which is completely eradicated from its resonance locations; the eccentricities increase until they cross the planets\rev{'} orbits and scatter. \rev{Some} planetesimals trapped in the 2:1 resonance, the closest to Planet b, survive until the end of the simulation. The 2:1 resonance also drives the highest eccentricity values in the trapped bodies\rev{, which drives some to close encounters with Planet b}.

\rev{The formation of the disc’s inner rim with a resonant population parallels the scenario of the Solar System involving Neptune and the Kuiper Belt \citep{Levison&Morbidelli2003}. Although the Kuiper Belt has a significant population at the 3:2 MMR with Neptune \citep{Gladman&Volk2021}, our simulation reveals only a weak accumulation at that location, with most bodies instead concentrating at the 2:1 MMR. The massive gravitational influence of Planet b would likely lead to a concentration at a more-distant MMR due to its more extensive chaotic zone, which, as discussed, ultimately removes particles trapped at closer MMRs (e.g. \citealt{Pearce+2024})}.

Beyond the planetesimal disc's inner rim, the influence of high-order resonances is also detectable, particularly the 3:1 resonance, which plays a significant role in exciting the disc locally. Unlike the 2:1 resonance, high-order resonances do not trap a large number of bodies, which leads to the absence of gap regions in the debris disc. Nevertheless, they can excite planetesimals during their sweeping transit through the disc; for example, the 3:1 resonance excites the area between approximately 150 au and 170 au. By the end of the simulation, planetesimals remaining near or within resonant locations exhibit eccentricities below 0.2.

\begin{figure}
\centering
\begin{center}
    \includegraphics[width=0.45\textwidth]{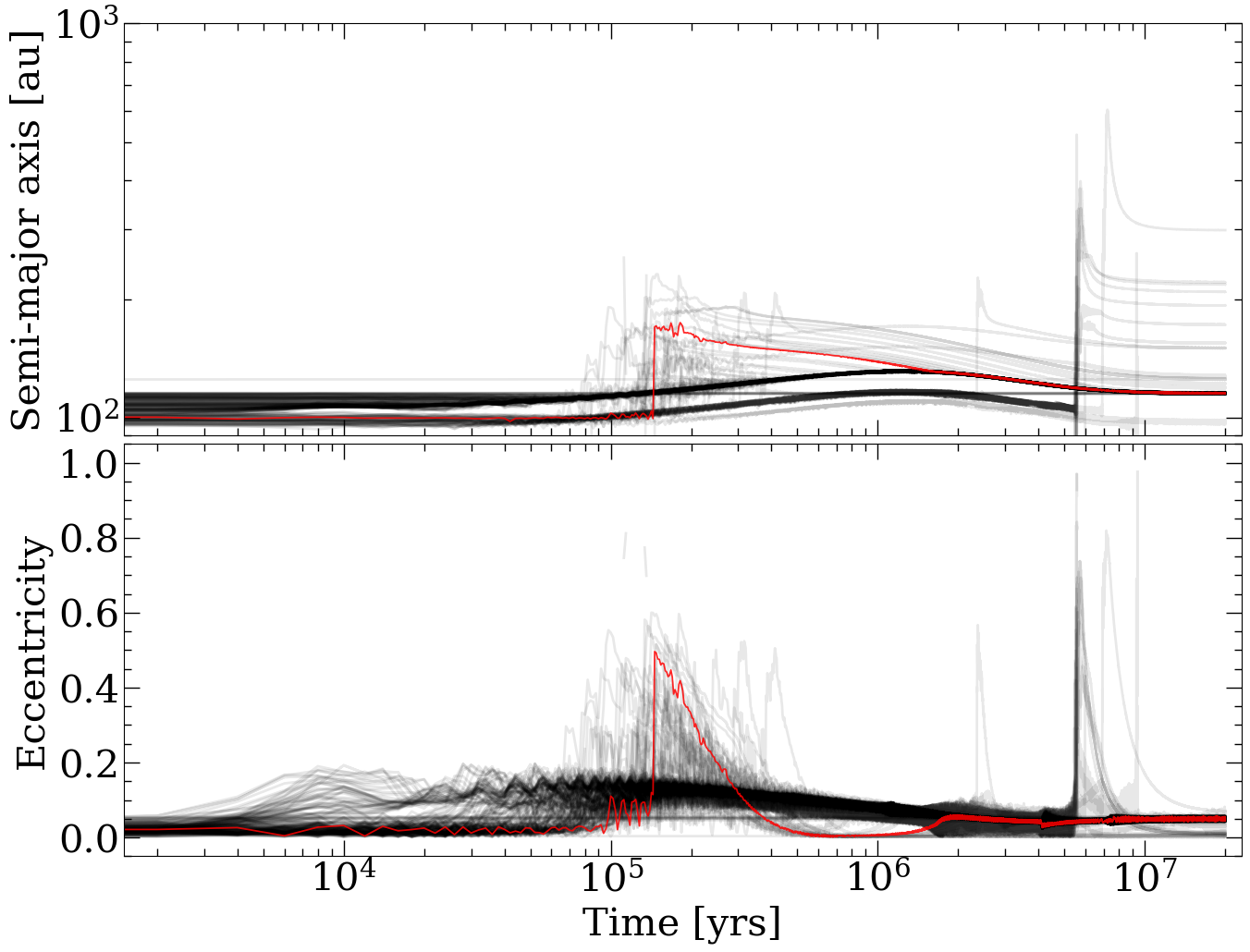} 
    \caption{Evolution of the semi-major axis and eccentricity \rev{of Planet x in the five-planet model,} for varying initial conditions. Gray lines indicate 200 simulations \rev{employing the {\sc IAS15} integrator} with different initial values for \rev{Planet x's} semi-major axis and true anomaly. The red line represents the case displayed in Figures \ref{fig:main_plot_5p} and \ref{fig:main_plot_5p_2}.}
    \label{fig:chaos_x}
\end{center}
\end{figure} 

The surface-density profile at the inner rim can be described by an erf~function \citep{Rafikov2023}. We fit the following profile to the whole disc, using the method in \citet{Pearce+2024}:
\begin{equation}\label{eq:SDprofile}
    \Sigma(r) = \frac{\Sigma_0}{2} \left[1-\text{erf}\left(\frac{r_{\rm in}-r}{\sqrt{2}\sigma_{\rm in}r_{\rm in}}\right)\right]\left[1-\text{erf}\left(\frac{r_{\rm out}-r}{\sqrt{2}\sigma_{\rm out}r_{\rm out}}\right)\right]\left(\frac{r}{r_{\rm in}}\right)^{-\gamma},
\end{equation}
where $\Sigma_0$ is the surface density at $r_{\rm in}$ which is the location of the inner edge of the disc, meanwhile $\sigma_{\rm in}$ is a measure of the edge steepness. $r_{\rm out}$ and $\sigma_{\rm out}$ refer to the outer edge measurements, and $\gamma$ represents the power law index. The resulting fitting profile is shown in the lower row of Figure \ref{fig:main_plot_2} and the obtained values for the inner edge region are listed in Table \ref{tab:4planetsSD}. We find that the obtained values are low to correctly reproduce the HR~8799 debris disc measurements even when a smoothing function is applied to the radial profile.

Firstly, both values $r_{\rm in}$ and $\tilde{r}_{\rm in}$ for the raw and smoothed data, respectively, are lower than the observed value by the end of the simulation. This is primarily due to a significant number of bodies still being located within 145 au, as illustrated in Figure \ref{fig:main_plot_2}. Based on this, we find that a four-planet architecture is unable to produce a sufficiently large cavity size, even when considering the outward migration scenario. Nevertheless, it can create a larger cavity than that seen in non-migrating scenarios \citep{Read+2018}. 

Secondly, the final $\sigma_{\rm in}$ value obtained, $0.04$, is also low compared to the observed value. This value is associated with the eccentricity of the planetesimals; a lower eccentricity results in a steeper edge and, hence, a smaller sigma value. However, the value is also sensitive to the smoothing effect of applying a filter or beam. This effect becomes evident when examining the $\tilde{\sigma}_{\rm in}$ value, derived by applying a Gaussian beam size of 43 au to the raw data; the same beam size as \citet{Faramaz+2021} employed on their data. Although a beam effect can produce high $\sigma_{\rm in}$ values, the observed value better aligns with the presence of a scattered population within the disc (see Section \ref{sec:HR}). In our four-planet simulation, the highest $\sigma_{\rm in}$ value is obtained at 1.0 Myr, corresponding to the initial stages where the excitation level is primarily influenced by Planet b. By the end of the simulation, the number of highly excited planetesimals decreases, leading to a corresponding decline in the $\sigma_{\rm in}$ value.

The inconsistencies noted in the edge size and shape of the inner disc raise questions about the feasibility of the current four-planet model. \rev{This edge size issue is also observed in $\beta$~Pictoris between its debris disc and its two detected planets; 
\citet{Lacquement+2025} examined the possibility that additional planets could carve a larger cavity, which would align with the observations.} This prompts us to investigate \rev{a similar} scenario in which we incorporate an extra body to determine whether these changes could replicate the observed characteristics better. 



\begin{figure*}
\centering
\begin{tabular}{cc}
    \includegraphics[width=0.45\textwidth]{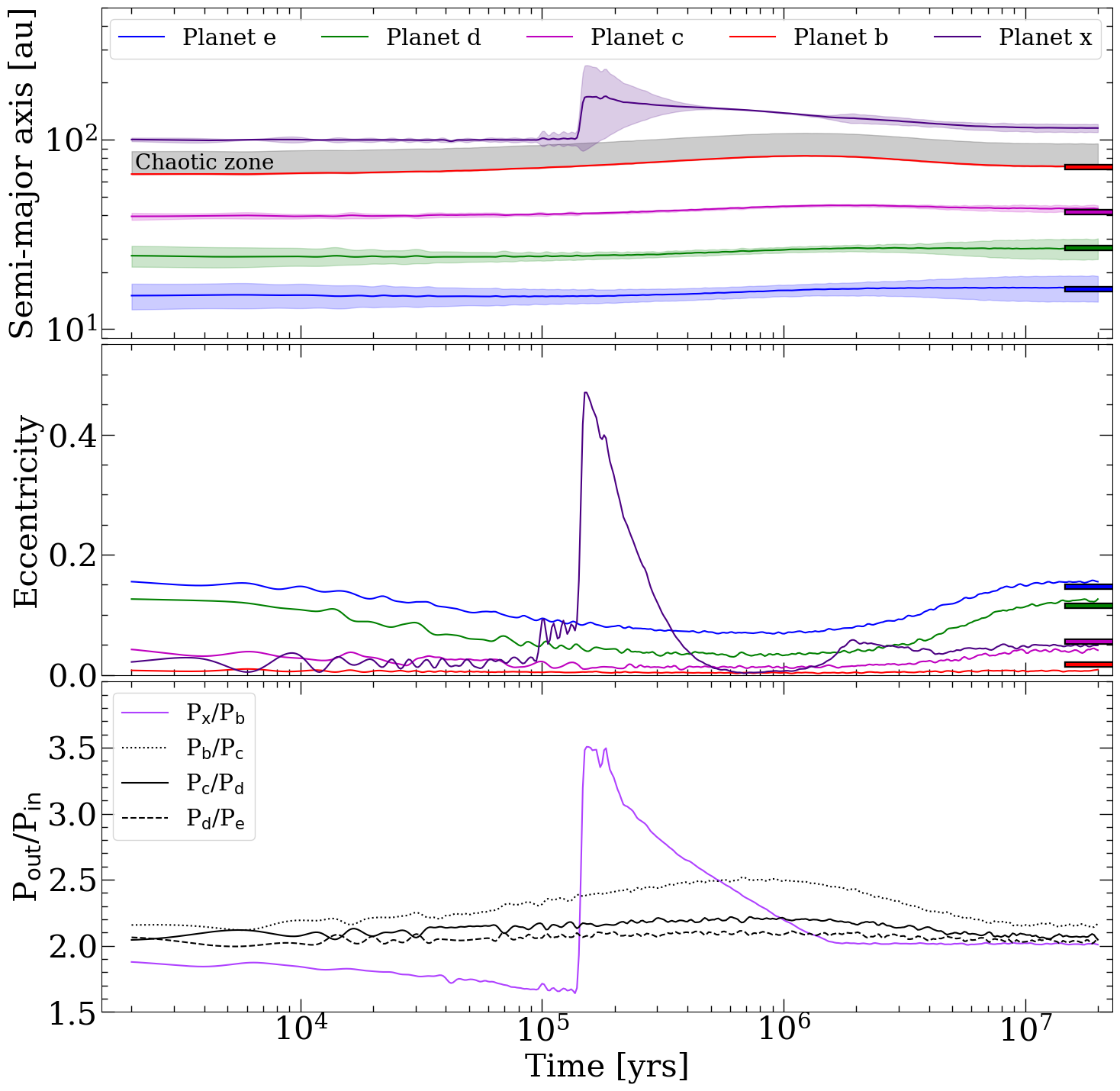}&
    \includegraphics[width=0.46\textwidth]{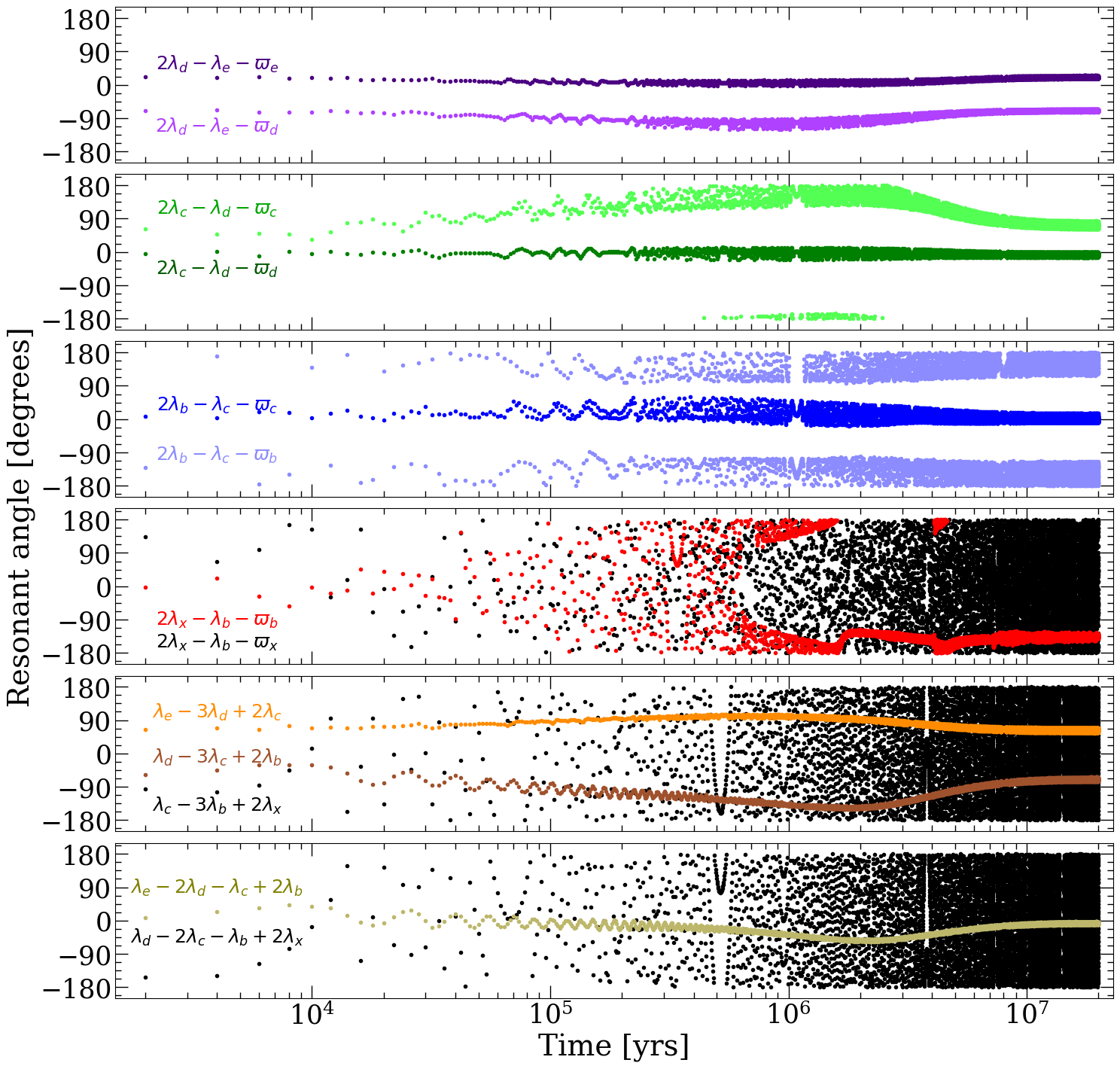} \\
\end{tabular}
    \caption{Evolution of selected orbital parameters \rev{in the five-planet model}. \textbf{Left}: The evolution of the planets' semi-major axes, eccentricities, and the period ratio of \rev{adjacent-planet pairs}. Results for planets e, d, c, b, and x are depicted by using the same color-coding as in Figure \ref{fig:sketch} plus violet for Planet x. \rev{The gray region indicates the outer chaotic zone for Planet b, defined as 3 Hill radii.} \textbf{Right}: The resonant angles for each adjacent pair, trio, and quartet of planets are shown.}
    \label{fig:main_plot_5p}
\end{figure*} 

\begin{figure*}
\centering
    \includegraphics[width=1.0\textwidth]{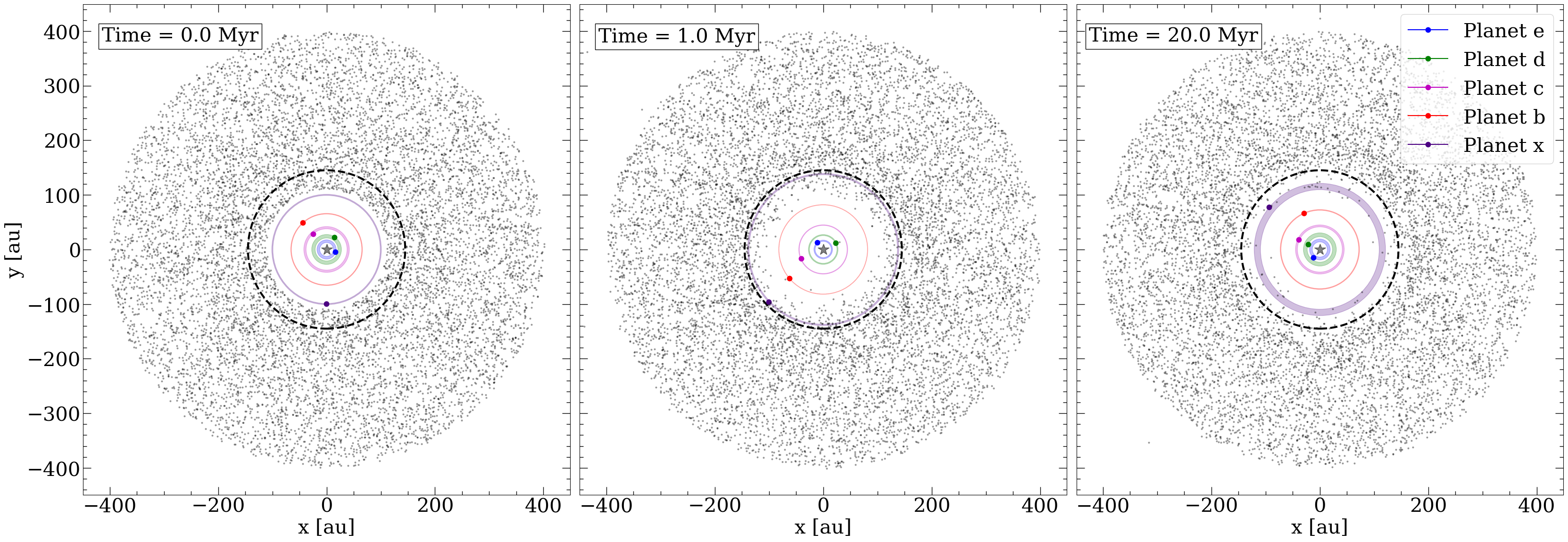}\\ \includegraphics[width=1.0\textwidth]{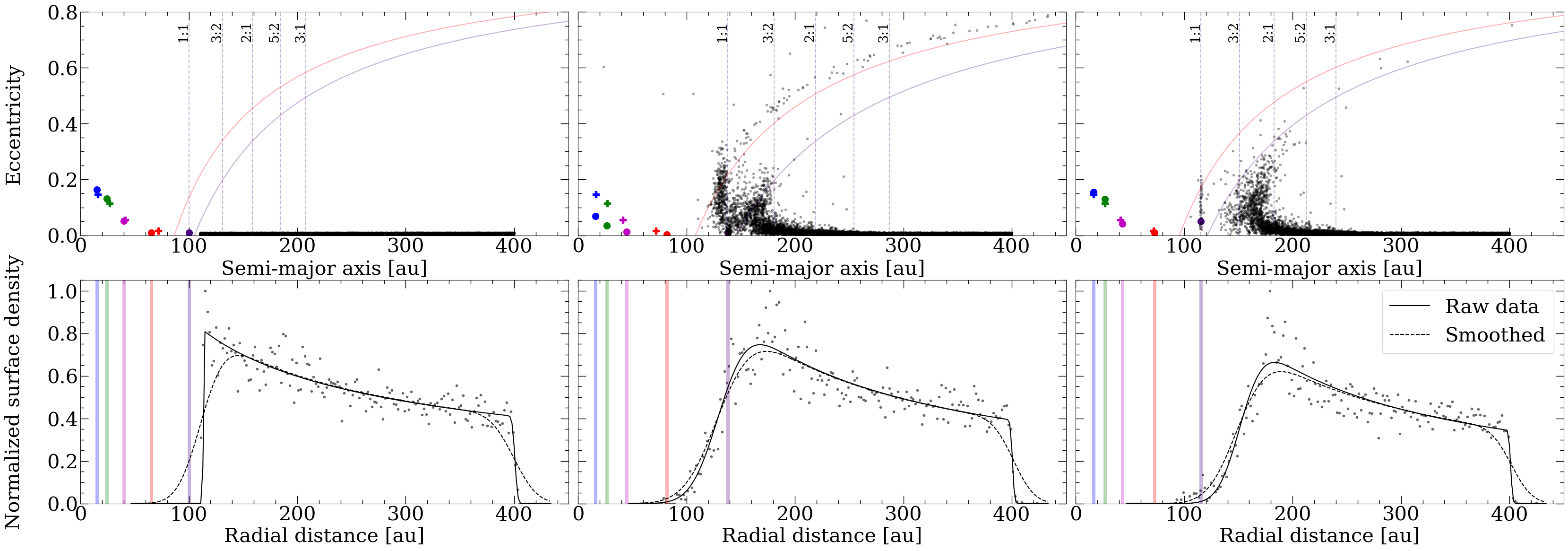} \\
    \caption{Same as Figure \ref{fig:main_plot_2}, but adding an extra planet. The MMR lines are plotted with respect to Planet x. }
    \label{fig:main_plot_5p_2}
\end{figure*} 

\subsection{Fiducial model - five planets}
\label{sec:five-planets scenario}

\rev{We incorporate a fifth planet, referred to as Planet x, which is initially positioned in a non-resonant orbit with an inclination of $0.5\degree$ and eccentricity of 0.01. This configuration is selected because there is no evidence for a globally eccentric outer debris disc (see Section \ref{sec:Architecture}) that would suggest an initially excited configuration. Planet x is initialized at 100 au,} close to the 2:1 resonance location relative to Planet b. These assumptions contribute to a rapid orbital kick of Planet x due to dynamical orbital instabilities triggered by Planet b. While this effect can occur in other locations of MMR, we chose the 2:1 resonant location to ensure consistency with the overall planetary architecture. The considered mass is set to 0.025 $M_{\rm Jup}$ ($\approx 8\ M_{\oplus}$) which stays below the detection threshold. It is near the mass limit suggested by \citet{Read+2018} to reproduce the radial profile without disrupting the stability of the other planets or causing a loss of the MMR. We enforce that Planet x undergoes inward migration similar to Planet b, with a migration timescale of $\tau_{a,x} = 8$ Myr. Additionally, the initial inner edge of the planetesimal disc is set further at 110 au. The rest of the disc and planets parameters are the same as listed in Table \ref{tab:planets_params}, the fiducial model.

Due to the expected orbital instability arising from the interaction between Planet b and Planet x, the dynamical behavior of Planet x is anticipated to be chaotic. To assess possible outcomes, we ran 200 simulations featuring only the five planets, varying Planet x's semi-major axis and true anomaly randomly within the ranges of 95-115 au and 0-360$\degree$, respectively. The range in the semi-major axis spans the region between the 3:2 and 2:1 MMR locations with Planet b. Figure \ref{fig:chaos_x} shows the evolution of these quantities. The simulations indicate a preferred evolutionary pathway for Planet x: almost 70\% of the simulations follow the pathway represented by the two thickest lines at roughly 1 Myr, increasing to 80\% by the end, which corresponds to a close location with the 2:1 MMR of Planet b. We employed the setup highlighted in red, and although it represents roughly 10\% of the simulations' starting behavior (where their eccentricities increase abruptly), it maximizes the number of scattered bodies during the early stages, but ends \rev{up similar to} most of the simulations.

Figures \ref{fig:main_plot_5p} and \ref{fig:main_plot_5p_2} illustrate the evolutionary state of a planetary system and its planetesimal disc across three epochs for a scenario consisting of five planets. We find that the additional planet does not affect significantly the evolution of the four innermost planets compared to the fiducial model. 
\rev{The gradual outward migration of Planet b and the inward motion of Planet x bring their orbits closer together. This convergence continues until Planet x enters the chaotic zone of Planet b, leading to dynamical instability. Around 0.13 Myr, this interaction triggers a sudden jump in both the semi-major axis and eccentricity of Planet x. As a result, the fifth planet is displaced to a larger radial distance, embedding it within the planetesimal disc.} After this displacement, Planet x continues its inward migration and eventually becomes trapped in a 2:1 resonance with Planet b, without forming a multi-resonant state with the other planets.
\revII{Note that in order to have a resonance between two planets, at least one of the resonant angles needs to librate, not both. In our simulations, even when the two planet angles $\phi_1 = 2\lambda_b - \lambda_c - \varpi_{b,c}$ and $\phi_{2,b} = 2\lambda_x - \lambda_b - \varpi_b$ (see Figure \ref{fig:main_plot_5p} right panels) exhibit bounded libration, $\phi_{2,x} = 2\lambda_x - \lambda_b - \varpi_x$ does not. For this reason, their difference $\phi_2 - \phi_1 = \lambda_c - 3\lambda_b + 2\lambda_x$ involving the three planets does not librate. Consequently, the four-planet angle involving Planet x oscillates as well.}

The jump in radial distance of Planet x significantly impacts the planetesimal disc, as it moves inward.
This leads to the formation of a population of co-rotational planetesimals in the early stages of evolution and noticeably excites surrounding bodies. The excited bodies reach distances up to 250 au, a range not reached in the four-planet simulation. The highly eccentric bodies interact with Planet b as well as Planet x, resulting in outcomes similar to the four-planet simulation, where they cross the orbits of interior planets and are subsequently ejected. Ultimately, a considerably larger number of high-eccentricity planetesimals survive, delineating the boundary in the $ae$-plane equivalent to 3 Hill radii of Planet x, consistent with expectations for a scattered population \citep{Gladman1993,Ida+2000b,Ida+2000a, Pearce+2022}.

The resulting values of $r_{\rm in}$ and $\sigma_{\rm in}$ in this case better explain the observations than the four-planet case scenario listed in Table \ref{tab:4planetsSD}. By the end of the simulation, with Planet x at approximately 120 au, the $r_{\rm in}$ value aligns with observations at 160 au. Despite $\sigma_{\rm in}$ increasing by almost twice when compared to the four-planet case, which implies the huge number of high-eccentric bodies, the value still remains insufficient\rev{; our simulated-disc edge is still steeper than in the observed disc}.

\subsection{On surface density radial profile} \label{sec:sim_obs}

The radial profile presented in \citet{Faramaz+2021} reveals an asymmetric and non-sharp profile at the inner edge between 100 and 200 au. The shape of the inner edge is consistent with the existence of a high-eccentricity component or a non-sharp cutoff in the distribution of semi-major axes \citep{Marino2021,Pearce+2024}.

\begin{figure}
\centering
\begin{center}
    \includegraphics[width=0.45\textwidth]{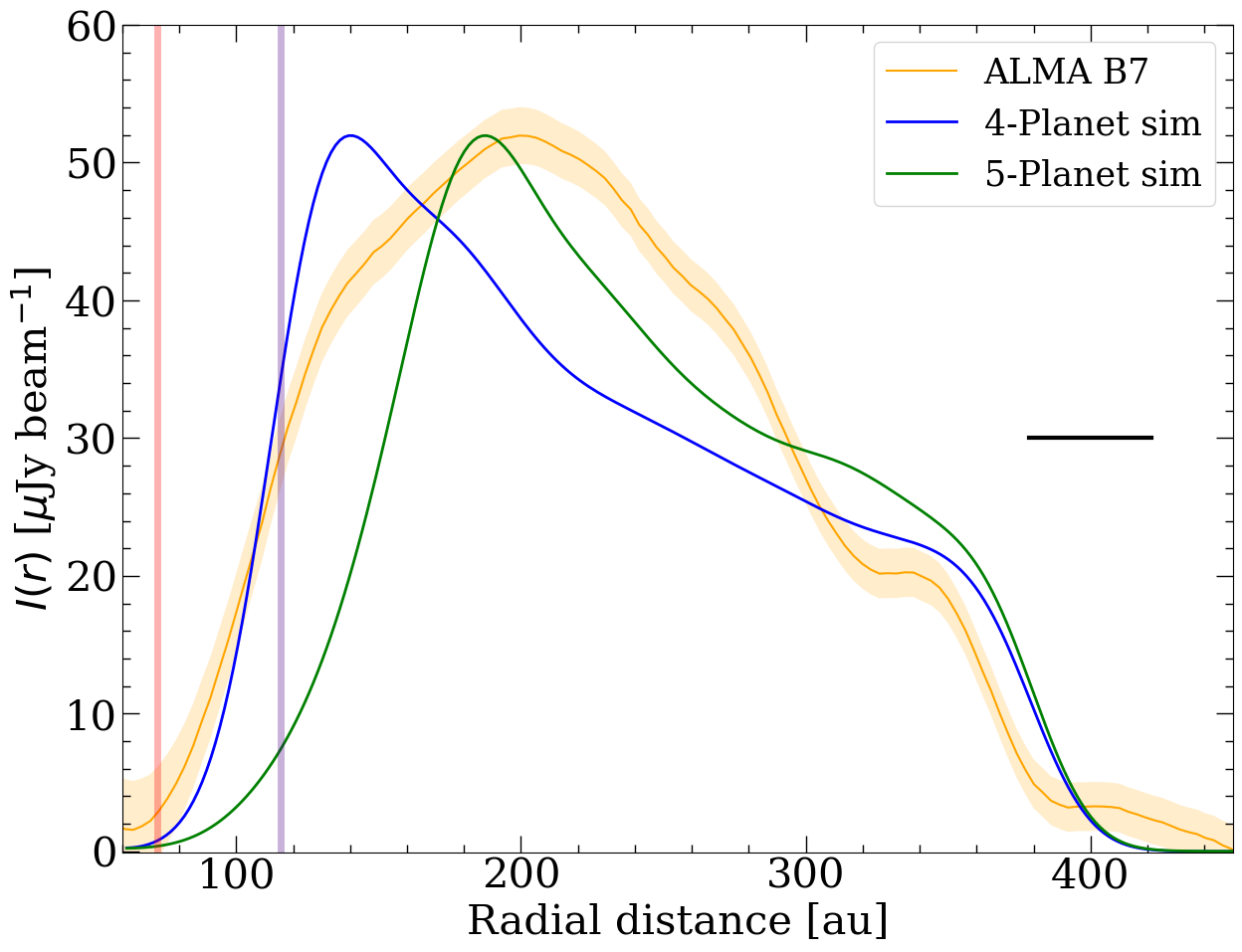} 
    \caption{Radial intensity for the ALMA observation (orange line) along with our two relative intensity profiles for the four- and five-planet simulation in blue and green, respectively. The simulation relative intensity is calculated by assuming the planetesimals emit as blackbodies, following the relationship $I(r) \propto \Sigma(r) r^{-1/2}$. The intensity is smoothed using a Gaussian filter of 43 au. The beam size appears as the black line. The locations of Planet b (red) and x (violet) are marked by vertical lines for reference.}
    \label{fig:ALMA+sim}
\end{center}
\end{figure}

Figure \ref{fig:ALMA+sim} compares the ALMA Band 7 observation with our simulation results. In contrast to the radial profiles produced by the four- and five-planet simulations, the observed profile appears broader and flatter around its peak intensity at about 200 au. The four-planet simulation does not replicate the shape and steepness of the inner edge, even after accounting for planetary migration. While the five-planet profile effectively matches the peak intensity location and the size of the inner edge, it still fails to capture the steepness and shape seen in the observed data. This suggests that additional factors may influence the surface density profile of HR~8799, which are not accounted for in our current models. 

The complexity of the disc profile suggests that various mechanisms may be at play, capable of generating and sustaining a significant number of eccentric bodies. One such mechanism is the influence of a dwarf planet embedded within the planetesimal disc \citep{Ida+1993,Kirsh+2009,Friebe+2022,Costa+2024}. A dwarf planet (or planets) would effectively account for the high levels of eccentricity within the disc, acting similarly to Pluto in our Solar System by exciting nearby planetesimals without significantly clearing its own orbit \citep[see][for a review on transneptunian objects]{Gladman&Volk2021}. This scenario also aligns with proposed Pluto formation models that involves Neptune's outward migration phase \citep{Malhotra1993} similarly to our migration model. Self-stirring is another potential source of eccentricity excitation, resulting from the self-gravity of bodies within the debris disc \citep{Kennedy&Wyatt2010,Krivov+2018,Poblete+2023}. \rev{A third possibility is collisions between planetesimals, which would cause a less-steep edge profile \citep{ImazBlanco+2023}. The combined effect of collisions plus planetary sculpting would produce a smoother, `kinked' edge compared to planets alone, which appears similar to that of the observed disc (\mbox{Figure 13} in \citealt{Pearce+2024}).}

\section{Discussion}
\label{sec:discussion}

Tracing the dynamical history of a planetary system is complicated, even for our Solar System. As we have seen in Section \ref{sec:HR}, the HR~8799 system has significant pieces of information to try to trace its past. Our model uses the constraints available and explores an evolutionary pathway apt for super-Jupiter planets in a resonant chain. The parameterization of the migration regimes provides a reasonably good approach to modeling the system; nevertheless, both the model and observations have essential aspects to highlight and discuss.    


One of our key hypotheses is that the four super-Jupiters formed rapidly and efficiently. This assumption is critical for initiating the outward migration during the early stages of planetary formation when enough gas is present to drive Planet b inward. Therefore, a swift planetary system formation is necessary, along with an efficient mechanism to trap consecutive pairs of planets into a 2:1 MMR and sufficient time to allow the subsequent outward and inward migrations.

Recent observations suggest that the gap-opening process associated with massive planet formation can commence as early as approximately 0.5 Myr \citep{Segura-Cox+2020}, and potentially even at ages less than 0.2 Myr \citep{Maureira+2024}. This implies that at least one Jupiter-like planet is already present at such an early stage. \rev{\citet{Lambrechts&Johansen2012} estimated the time needed to reach the critical core mass (10 $M_\oplus$) via pebble accretion, which is necessary for starting gas accretion. By applying our radial locations ($a_{\rm init}$) and surface density values, we determine the timescales to be 0.24, 0.23, 0.25, and 0.29 Myr for Planets e, d, c, and b, respectively, all of which are less than 0.5 Myr. Notably, Planet d would form first, potentially explaining its higher mass as suggested by models (see Section \ref{sec:Architecture}). Additionally, the hypothetical Planet x would require 0.34 Myr to achieve a mass of 8 $M_\oplus$.}

\citet{Lau+2024} have shown a sequential inside-out planet formation mechanism, whereby the formation of one gap-opening planet creates pressure maxima at the edge, triggering the formation of an adjacent planet. This process also results in the planets trapped in a 2:1 resonant chain, supporting part of our hypothesis's initial conditions. Such a mechanism primarily investigates the formation of gas and ice giants without considering larger planets, but more massive planets were not considered.

Regarding planetary growth, \citet{Lambrechts+2019} estimates that the accretion rate of planets could reach values as high as 10 $M_{\rm jup}$/Myr. Conversely, the minimum gas mass required to form the observed planets in HR~8799 is approximately 0.03 $M_{\odot}$, likely concentrated within the inner 100 au. The size of the debris disc suggests an initial disc extension of around 400 au or even more. Consequently, the total initial disc mass would be significantly larger, and according to the disc-to-star mass ratios deduced by \citet{Andrews+2013}, it could approach 0.1 $M_{\odot}$. Fast gas accretion, supported by a substantial mass reservoir, could facilitate the rapid growth of protoplanets, reaching their current masses within the first million years.

The timing of the onset of outward migration for the three inner planets is also crucial for assessing the efficiency of the process. As explained in Section \ref{sec:mech+migration}, outward migration is driven by the formation of an eccentric gap, which typically occurs for planetary masses larger than 3 $M_{\rm jup}$. The values we used to compute migration timescales align with the present measurements for the planets' masses, all of which are greater than 6 $M_{\rm jup}$. The calculated dynamical masses of the HR~8799 planets can vary depending on the star model used; however, most models concur that all planetary masses are similar, except for the outermost planet, which is less massive than the rest. This implies that the three inner planets may have formed almost simultaneously, while the outermost planet could have formed slightly later \citep{Bergez-Casalou+2023}. Consequently, the outward migration of the three inner planets might have begun during their growth phase, occurring relatively simultaneously. This results in a distinct set of parameters for the three planets migrating outward. Moreover, the $\tau_{a,i}$ and $\tau_{a,i}^+$ values are derived from single-planet simulations; introducing a pair of planets could alter the total torques, affecting the timescale and/or direction of migration. 

The eccentricity imprinted on the disc by the eccentric gaps would modify the torques experienced by the planetesimals (which are initially assumed to be circular and massless in our model), potentially altering their migration timescales and leading to a different final architecture and morphology of the debris disc than we observed \citep[see eccentric debris discs in][for instance]{Kennedy2020}. While the super-Jupiter planets are too massive to be significantly influenced by planetesimal-induced migration, this effect could be significant for a potential fifth planet (Planet x) and any smaller bodies, including dwarf planets, embedded within the planetesimal disc \citep{Kirsh+2009,Ormel+2012, Friebe+2022}. These smaller bodies may then have played an important role in sculpting the observed structure of the HR~8799 debris disc after the gas disc disappeared.


\section{Conclusion}
\label{sec:conclusion}

We performed self-consistent $N$-body simulations of a planetary system consisting of four super-Jupiter planets under both outward and inward migration regimes, as well as a planetesimal disc with massless particles to explain the HR~8799 architecture for both planets and the cold debris disc. Our main findings are summarized as follows:

\begin{enumerate}[(i)]

    \item Stability and robustness of resonance: The resonant configuration among the four planets can be maintained during outward migration, provided that the migration rate is not excessive. However, if the outward migration is too fast or if only the outermost planet migrates inward, the resonant structure can be disrupted, ultimately destabilizing the planetary architecture. 

    \item Disc structure and the role of an additional planet: Outward migration of the observed four planets significantly excites the planetesimal disc, leading to a transient scattered population and expanding the inner cavity beyond what models without migration predict. However, this alone does not fully reproduce the observed debris disc. Our results indicate that the presence of a fifth planet beyond Planet b’s orbit improves the agreement with observations, as it undergoes an initial instability before settling into a wider orbit. This additional planet helps shape the debris disc, suggesting that unseen planetary companions may play a key role in sculpting the system’s architecture.

    
        
\end{enumerate}

This work proposes a reinterpretation of the HR~8799 system's planetary architecture and disc morphology, incorporating a phase of outward planetary migration. While this model provides a partial explanation for several key features, it relies on specific initial conditions\rev{, migration regimes} and parameter choices, highlighting the need for further investigation into the broader range of dynamical processes that may influence planetary system evolution. The complex interplay between inward and outward migration, along with resonant interactions, emphasizes the limitations of our current understanding and the necessity for more comprehensive models encompassing a wider variety of initial conditions and dynamic processes.


\begin{acknowledgements} 
\rev{The authors thank the anonymous reviewer for the thorough and useful report.} We acknowledge Antranik A. Sefilian for his valuable help and support throughout this work. The authors also thank Virginie Faramaz for providing the HR~8799 ALMA data. P.P.P. is supported by the European Research Council (ERC) under the European Union Horizon Europe programme (grant agreement No. 101042275, project Stellar-MADE). T.D.P. is supported by UKRI/EPSRC through a Stephen Hawking Fellowship. 
C.C. is supported by ANID through FONDECYT grant No. 3230283. 
\end{acknowledgements}

\section*{Data availability}
The corresponding author will share the data underlying this article at a reasonable request. We used the following public softwares:\\ 
\newline
\begin{tabular}{ll}
    \textsc{Rebound } & \hyperlink{https://github.com/hannorein/rebound}{https://github.com/hannorein/rebound} \\
     &  \citet{rebound} \\
     \textsc{Reboundx} & \hyperlink{https://github.com/dtamayo/reboundx}{https://github.com/dtamayo/reboundx} \\
     &  \citet{reboundx}
\end{tabular}

\bibliographystyle{aa}
\bibliography{paper}




\label{lastpage}
\end{document}